\documentclass[11pt,a4paper]{article} 
 	
        \usepackage[utf8]{inputenc}
        \usepackage{fontenc}
	\usepackage[italian,english]{babel}
	\usepackage{amsmath}
	\usepackage{geometry}
	\usepackage{amsthm}
	\usepackage{mathrsfs}
	\usepackage{bbold}
	\usepackage{mathtools}
	\usepackage{amsfonts}
	\usepackage{graphicx}
	\usepackage{physics}
	 \usepackage{comment}
	 \usepackage{bbm}
	  \usepackage{layout}
	\usepackage{xcolor}
     \usepackage[big]{layaureo}
	 \usepackage{dsfont}	
	 \usepackage{slashed}
 	 \usepackage{amssymb}
	 \usepackage{tensor}
	\usepackage{fancyhdr} 
 	\usepackage{tikz-cd}
 	\usepackage{braket}
 	\usepackage{tikz}
 	\usepackage[compat=1.1.0]{tikz-feynman}
 	\usepackage{subfigure}
	\usepackage{fancyhdr} 
	\usepackage{booktabs}
	\usepackage{braket}
	
\renewcommand{\i}{\mathrm{i}}

\usepackage{tikz}

\usetikzlibrary{
	decorations.markings,
	decorations.pathmorphing,
}

\tikzset{cutstyle/.style={decorate, decoration={zigzag, segment length=6, amplitude=2}, draw=black}}
\tikzset{arrow data/.style 2 args={decoration={markings, mark=at position #1 with \arrow{#2}}, postaction=decorate}}
\tikzset{partial ellipse/.style args={#1:#2:#3}{insert path={+ (#1:#3) arc (#1:#2:#3)}}}

\usepackage{jheppub} 

\usepackage{natbib}
	\newcommand{\beq}{\begin{equation}}
	\newcommand{\bea}{\begin{eqnarray}}
	\newcommand{\eea}{\end{eqnarray}}
	\newcommand{\eeq}{\end{equation}}

	\makeatletter
	\newcommand{\aextp}{\@ifnextchar^\@aextp{\@aextp^{\,}}}
	\def\@aextp^#1{\mathop{\bigwedge\nolimits^{\!#1}}}
	\makeatother
	
	\makeatletter
	\newcommand{\extp}{\@ifnextchar_\@extp{\@extp_{\,}}}
	\def\@extp_#1{\mathop{\aextp\nolimits_{\!#1}}}
	\makeatother

	\theoremstyle{definition}

\title{Sine-dilaton gravity vs double-scaled SYK: exploring one-loop quantum corrections}

\author[a]{Leonardo Bossi,}
\author[a]{Luca Griguolo,}
\author[b]{Jacopo Papalini,}
\author[c]{Lorenzo Russo,}
\author[c]{and Domenico Seminara}

\affiliation[a]{Dipartimento SMFI, Universit\`a di Parma and INFN Gruppo Collegato di Parma,
	Viale G.P. Usberti 7/A, 43100 Parma, Italy}
\affiliation[b]{Department of Physics and Astronomy
Ghent University, Krijgslaan, 281-S9, 9000 Gent, Belgium}
\affiliation[c]{Dipartimento di Fisica, Universit\`a di Firenze and INFN Sezione di Firenze, via G. Sansone 1, 50019 Sesto Fiorentino, Italy} 

\emailAdd{leonardo.bossi@unipr.it}
\emailAdd{luca.griguolo@unipr.it}
\emailAdd{jacopo.papalini@ugent.be}
\emailAdd{lorenzo.russo@unifi.it}
\emailAdd{domenico.seminara@unifi.it}

\abstract{We provide non-trivial checks of the recently proposed duality between double-scaled SYK and a 2d dilaton gravity model with sine potential, studying the path integral at one-loop level. Specifically, we compute the logarithmic correction to the free energy of sine-dilaton gravity and, up to potential ordering ambiguities, we find a match with the corresponding quantity in double-scaled SYK. The computation relies on the description of sine-dilaton gravity in terms of a version of the q-Schwarzian theory, the quantum deformation of the standard Schwarzian model dual to JT gravity. A crucial aspect of the calculation is selecting the correct Hartle-Hawking vacuum for the gravitational theory, which implies a specific choice of boundary conditions for the one-loop determinant, computed using a generalization of the Gel'fand-Yaglom's theorem. We also evaluate the gravitational one-loop correction to the boundary to boundary propagator of a non-minimally coupled matter field in the bulk theory, showing a perfect agreement with the corresponding quantum correction of matter correlators in double-scaled SYK.    }

\begin{document}
\maketitle
\flushbottom

\section{Introduction}

The Sachdev-Ye-Kitaev (SYK) model \cite{Sachdev_1993, Sachdev:2010um,kitaev2015simple, Maldacena:2016hyu} is a quantum mechanical system comprising \( N \) Majorana fermions interacting through random all-to-all \( p \)-local interactions. Historically, it was first introduced\footnote{Earlier incarnations in nuclear physics appeared in \cite{FRENCH1970449, BOHIGAS1971261}} in condensed matter physics as a model for strange metals \cite{Sachdev_1993, Sachdev:2010um, Chowdhury:2021qpy}. More recently, it has garnered significant attention within the holographic community \cite{kitaev2015simple,Maldacena:2016upp, Cotler:2016fpe, Saad:2018bqo, Maldacena:2018lmt, Goel:2018ubv, Jensen:2016pah, Polchinski:2016xgd, Kourkoulou:2017zaj}. The SYK model serves as a toy model for testing quantum gravity, representing one of the rare cases that is both analytically tractable in the infrared (IR) and other regimes while also featuring a maximal chaos exponent.

More specifically, in the IR, the model exhibits a conformal regime with a pattern of symmetries encoded by Schwarzian quantum mechanics, which governs the effective dynamics of gravity on \(\text{AdS}_2\) coupled to a scalar field \cite{kitaev2015simple, Maldacena:2016hyu, Maldacena:2016upp, Sarosi:2017ykf, Berkooz:2018jqr}. These insights have been derived by analyzing the SYK model in the large-\( N \) limit, where the interaction order \( p \) is held fixed. Within this framework, one can formulate Schwinger-Dyson consistency equations for the two-point function \cite{Maldacena:2016hyu}, which are solvable in the IR using a conformal ansatz \cite{Maldacena:2016hyu, Maldacena:2016upp,  Mertens:2017mtv}. At low energies, an emergent reparametrization symmetry gives rise to Schwarzian theory as the effective low-energy description of the SYK model, establishing a duality with the well-known JT gravity \cite{Maldacena:2016upp, Engelsoy:2016xyb, Jensen:2016pah, Mertens:2022irh}.

This correspondence has advanced our understanding of quantum black holes and quantum gravity in \(\text{AdS}_2\). However, finding  the holographic bulk dual of the complete SYK model remains an open question. To move beyond the low-energy limit, a different large-\( N \) scaling has been proposed, offering a more comprehensive solution across all energy scales. In this approach, \( p \) is not held fixed as \( N \to \infty \); instead a new parameter \(\abs{\log q} = \frac{p^2}{N}\) \footnote{We follow here the convention of \cite{Blommaert:2023opb,Blommaert:2023wad,Blommaert:2024ydx}, where the definition of $q$ differs from that in \cite{Berkooz:2018jqr,Berkooz:2018qkz,Lin:2022rbf} by $q^2=q_\text{there}$. Hence $\lambda_\text{there}=2\abs{\log q}$. } is kept finite, defining the so-called double-scaled SYK (DSSYK) model \cite{Cotler:2016fpe, Berkooz:2018jqr, Berkooz:2018qkz, Berkooz:2024lgq}. The JT regime is recovered in the limit \(\abs{\log q}\to 0\), while focusing on low-energy dynamics.

Importantly, all DSSYK amplitudes have been calculated using Hamiltonian methods \cite{Berkooz:2018jqr, Berkooz:2018qkz,Berkooz:2024lgq}, revealing a remarkable structural similarity to the amplitudes computed for JT gravity \cite{Mertens:2017mtv, Yang:2018gdb, Mertens:2018fds, Blommaert:2018oro, Iliesiu:2019xuh}. This resemblance naturally raises the question of whether DSSYK admits a gravitational dual. Such a dual could provide a UV completion of JT gravity and possibly capture additional features of the full SYK model \cite{Maldacena:2016hyu, Lin:2022rbf, Das:2017pif, Das:2017hrt, Goel:2021wim, Blommaert:2023opb, Blommaert:2023wad, Blommaert:2024ydx}. A plausible direction is to identify a dual gravity model within the general class of two-dimensional dilaton gravity theories. 

A remarkable example is sinh-dilaton gravity, also known as Liouville gravity \cite{StanfordSeiberg, Fan:2021bwt, Mertens:2020hbs, Blommaert:2023wad, Kyono:2017pxs, Collier:2023cyw, Turiaci:2020fjj}, which can be viewed not only as a 2D quantum gravity theory but also as a non-critical string theory involving one timelike and one spacelike Liouville field. This model has garnered significant attention, particularly due to its string-theoretical construction \cite{Collier:2023cyw}, which extends the so-called minimal string framework \cite{Seiberg:2003nm, Seiberg:2004at,Mertens:2020hbs}. Its amplitudes can be expressed as worldsheet CFT correlators integrated over the moduli space of Riemann surfaces, and a nonperturbative formulation has been proposed using random matrix theory \cite{Collier:2023cyw, Johnson:2024bue, castro2024relation}.
On the field theory side, the model is exactly solvable. This solvability can be explained in terms of its symmetry structure, which is governed by a quantum group: the modular double of \( SL_q(2, \mathbb{R}) \) \cite{Mertens:2020hbs}.

Another noteworthy example is sine-dilaton gravity \cite{Blommaert:2024ydx,Blommaert:2023opb}, distinguished by a periodic dilaton potential. In this case, the symmetry structure is still governed by a quantum group, specifically \( SU_q(1,1) \). The model can also be formulated in terms of two Liouville fields with complex conjugate central charges \cite{Blommaert:2024ydx}, and a proposed string-theoretical formulation has recently been introduced \cite{Collier:2024kmo}. An intriguing feature of sine-dilaton gravity is that the effective local cosmological constant lacks a definite sign, thereby opening the possibility of describing de Sitter geometries within this framework. This aligns with the expectation that DSSYK is related to de Sitter physics, or in general to cosmological spacetimes \cite{Narovlansky:2023lfz, Verlinde:2024znh, Verlinde:2024zrh, Susskind:2022dfz, Susskind:2022bia, Susskind:2023hnj}.

The dilaton gravities discussed above share fundamental characteristics with their solvable linear counterpart, JT gravity. JT gravity can be expressed in a first-order formulation as a topological BF theory, with all its features encoded in the boundary dynamics. Its description involves a quantum particle on a non-compact group manifold, which is directly linked to the \( SL(2, \mathbb{R}) \) symmetry of the Schwarzian action. Similarly, sinh-dilaton and sine-dilaton gravities are also topological, as they can be reformulated as Poisson sigma models \cite{Ikeda:1993aj, Ikeda:1993fh, Schaller:1994es, Cattaneo:2001bp,  Mertens:2022irh, Blommaert:2023opb, Blommaert:2024ydx}. While their dynamics remain confined to the boundary, the group manifold is now described by a quantum group the modular double or \( SU_q(1,1) \) depending on whether the deformation parameter is real or complex. 

The energy in these systems corresponds to the Casimir operator of the relevant quantum group, whose classical limit aligns with the Hamiltonian derived from the quantum Schwarzian \cite{Blommaert:2023opb}. Notably, this Hamiltonian coincides with the DSSYK transfer matrix \cite{Berkooz:2018jqr, Lin:2022rbf}, suggesting a clear duality between the gravitational theory and the statistical model. Furthermore, the quantum group framework provides valuable insights into the boundary conditions required for the dual gravitational description \cite{Blommaert:2023opb}. 

By starting with a continuum dynamical system describing a particle on \( SU_q(1,1) \) and imposing appropriate constraints, one can derive a \( q \)-Schwarzian phase space path integral for a single asymptotic boundary, and a \( q \)-Liouville phase space path integral for a Cauchy slice with two asymptotic boundaries \cite{Blommaert:2023opb, Blommaert:2023wad}. The amplitudes of the resulting constrained quantum mechanical system are equivalent to those of DSSYK. 

The precise holographic duality between DSSYK and 2d sine-dilaton gravity has been explored more directly in \cite{Blommaert:2024ydx}. There, the ADM Hamiltonian of sine-dilaton gravity was reformulated as a \( q \)-Liouville model via a canonical transformation, similar to what was done in the JT case \cite{Harlow:2018tqv}. However, this matching is subtle and requires careful consideration of key elements in the gravity model, such as the distinction between fake and real temperatures, the choice of the Hartle-Hawking vacuum, and the imposition of non-trivial geometric constraints \cite{Blommaert:2024ydx}. 

This approach also allows for the study of correlation functions. For example, in the JT case, correlation functions of a scalar field in the bulk can be analyzed via their duality with conformal bilocal operators on the boundary \cite{Bagrets:2016cdf, Bagrets:2017pwq, Mertens:2017mtv}. In the BF formalism, this corresponds to evaluating the vacuum expectation values of a system of anchored Wilson lines. Similarly, for sine-dilaton gravity, one can introduce a suitably non-minimally coupled probe in the bulk \cite{Blommaert:2024ydx}, which corresponds to dual bilocal operators in the \( q \)-Schwarzian system.

The equivalence between the Hamiltonian of sine-dilaton gravity and the DSSYK transfer matrix naturally leads to the identification of the same amplitudes. However, it is important to investigate the path-integral formulation of the theory in detail, as it would offer a deeper understanding of the underlying gravitational dynamics. This approach was originally used in the context of JT gravity, where the path integral was performed over the Schwarzian mode, associated with the spontaneous breaking of conformal symmetry \cite{Maldacena:2016upp}.

While the canonical quantization picture of sine-dilaton gravity is well understood \cite{Blommaert:2024whf}, this paper provides a computation of both the partition function and the two-point correlators of sine-dilaton gravity, going beyond the leading semiclassical regime within the path integral framework. The motivation is twofold: first, to provide a non-trivial check of the proposed duality between DSSYK and the dilaton gravity model with a sine potential; and second, to gain a better understanding of the path integral formulation of sine-dilaton gravity. As we will see, our results can be interpreted as measuring the first quantum correction to the effective scalar curvature in sine-dilaton gravity and, through this, we will be able to test the reliability of the semiclassical weak-gravity expansion of the gravitational path integral for sine-dilaton gravity.

It is worth noting that the one-loop path integral over the Schwarzian mode in JT gravity already provides a complete answer, thanks to localization \cite{Stanford:2017thb, Griguolo:2023aem}. However, this is not the case here. For the correlators, the matching is even more complicated in standard JT gravity \cite{Griguolo:2021zsn, Lam:2018pvp, Haehl:2017pak, Qi:2019gny}, and going beyond the leading-order behavior is necessary to extract any physical meaning from the exact expressions. In our case, explicit computations could provide insight into more intricate aspects of the gravitational theory, such as the chaotic properties encoded in out-of-time-order correlators or the scattering of shock waves.

From a more formal perspective, one could also explore the contributions of non-trivial saddles (or instantons), if they exist. For instance, it should be possible to explain the structure of the physical spectrum. A similar effect was observed in JT gravity at finite cutoff, where the inclusion of non-perturbative saddles accounted for modifications to the spectral properties \cite{Griguolo:2021wgy}.

Concretely, we compute here the logarithmic correction to the free energy of sine-dilaton gravity and find that it agrees with the corresponding quantity in DSSYK, differing only by a simple numerical factor, which can be attributed to potential ordering ambiguities. A key technical aspect of the calculation involves the selection of the correct Hartle-Hawking vacuum for the gravitational theory, as this choice determines the boundary conditions for the one-loop determinant, that is evaluated using a non-trivial generalization of Gel'fand-Yaglom's theorem. Furthermore, we compute the one-loop correction to the boundary-to-boundary propagator of a non-minimally coupled matter field in the bulk theory, finding a perfect match with the corresponding quantum correction of matter correlators in DSSYK.

The structure of the paper is as follows: In Section \ref{back}, we provide the necessary background, reviewing the transfer matrix of double-scaled SYK, the duality between sine-dilaton gravity and DSSYK, and its semiclassical expansion. Section \ref{3} is dedicated to the one-loop computation of the partition function, starting from the path integral formulation. We discuss the technical details regarding the choice of boundary conditions and the evaluation of the relevant functional determinant. Section \ref{one} extends our analysis to the two-point correlation functions, showing a precise match with the corresponding result from the DSSYK model. Our conclusions and future directions are presented in Section \ref{5}. Additionally, we include two technical appendices: one for the computation of the Green functions used in the main text, and another for the Poisson sigma model formulation of sine-dilaton gravity.

\section{Background material}\label{back}
\subsection{The transfer matrix of double-scaled SYK}
In this section, we provide a brief overview of the SYK model and its double-scaled regime, where the system becomes exactly solvable via an auxiliary quantum mechanics, known as the transfer matrix of DSSYK \cite{Berkooz:2018jqr,Berkooz:2018qkz,Lin:2022rbf}. This overview is not aimed to be self-contained, but rather to introduce the essential elements needed to support the discussion of the gravitational dual in the following sections. For a exhaustive recent review on the topic, see \cite{Berkooz:2024lgq}.

The SYK model consists of a system of $N$ Majorana fermions \( \psi_i \) (\( i = 1, \dots, N \)) that satisfy the anticommutation relations \( \{ \psi_i, \psi_j \} = 2 \delta_{i,j} \). The dynamics are governed by an all-to-all \( p \)-body interaction, described by the Hamiltonian
\begin{equation}
H_{\mathrm{SYK}} = i^{p/2} \sum_{1 \leq i_1 < \cdots < i_p \leq N} J_{i_1 \cdots i_p} \psi_{i_1} \cdots \psi_{i_p} \,,
\end{equation}
where the couplings \( J_{i_1 \cdots i_p} \) are typically assumed to be Gaussian random variables.

In the double-scaling limit, \( N \) and \( p \) are taken to infinity, while keeping the ratio 
\begin{equation}
\abs{\log q}=\frac{p^2}{N}
\end{equation}
constant and finite. To explain the transfer-matrix approach to DSSYK, it is useful to focus on the moments 
\begin{equation}\label{moments}
m_k = \langle \text{tr} H^k \rangle_J
\end{equation}
of the partition function \( \langle \text{tr} e^{-\beta H} \rangle_J \), where $\langle \rangle_J$ represents the averaging over the random couplings.

The various moments can be visually represented using what are known as chord diagrams \cite{Berkooz:2018jqr,Berkooz:2018qkz,Lin:2022rbf}. To construct the chord diagram corresponding to the 
k$^{\mathrm{th}}$ moment, we begin by drawing a circle to represent the cyclic trace, and place $k$ nodes on it, each corresponding to an insertion of the Hamiltonian as in \eqref{moments}. We then connect pairs of nodes with chords, which arise from the application of Wick's theorem when we perform the  averaging over the random couplings. See for instance Figure 1 in \cite{Berkooz:2024lgq}. 

We finally need to sum over all possible Wick contractions, which corresponds to summing over all chord diagrams with $k$ nodes. 
In this sum, the value of each diagram is weighted by the number of chord intersections. Specifically, as determined in \cite{Berkooz:2018jqr,Berkooz:2018qkz}, each intersection contributes a factor of $q^2$ in the double scaled regime, and the total expectation value of the moment is given by
\begin{equation}\label{mom}
m_k = \sum_{\text{chord diagrams with } k \text{ nodes}} q^{2\text{\#}},
\end{equation}
where $\#$ stands for the total number of intersections associated with the diagram.

The sum in \eqref{mom} is generally difficult to compute, but one can handle it by iteratively constructing all chord diagrams using a transfer matrix \cite{Berkooz:2018jqr,Berkooz:2018qkz}. Starting from a point on the circle with no open chords, we move clockwise, opening or closing chords at each node. After $k$ steps, all chords must be closed. This method uniquely generates all possible chord diagrams \footnote{See for instance Figure 2 in \cite{Berkooz:2024lgq}}.
During the construction, if there are \( n \) open chords, opening a new chord adds no intersections and increases the number of open chords to \( n+1 \). When closing a chord, the number of intersections depends on its position in the stack, contributing factors of \( 1, q^2, \dots, q^{2n-2} \), according to \eqref{mom}. Summing over all possible diagrams gives a factor of \( \frac{1-q^{2n}}{1-q^2} \) for each chord closure. We can thus define the creation and annihilation operators, \(\alpha^{\dagger}\) and \(\alpha\), as well as the chord number operator \(\hat{n}\), through their action on the chord Hilbert space with basis states \(\ket{n}\):
\begin{equation}\label{op}
\hat{\alpha} \ket{n} = \ket{n-1}, \qquad \hat{\alpha}^{\dagger} \ket{n} = \ket{n+1}, \qquad \hat{n} \ket{n} = n \ket{n}.
\end{equation}
The transfer matrix can now be expressed in terms of these operators as \footnote{The normalization of $\hat{T}$ in left hand side of \eqref{transfer} is standard. See for instance \cite{Lin:2022rbf}. }:
\begin{equation}\label{transfer}
\sqrt{2 \abs{\log q}}\hat{T} = \hat{\alpha}^{\dagger} + \hat{\alpha} \frac{1 - q^{2 \hat{n}}}{1 - q^2}.
\end{equation}

The moments are computed as the transition matrix elements of the \(k^{\mathrm{th}}\) power of the transfer matrix between the initial and final states with no chords, i.e.
$
m_k = \langle 0 | \hat{T}^k | 0 \rangle$.
Finally, summing over the moments, the partition function of double-scaled SYK is given by:
\begin{equation}\label{part}
Z_{\mathrm{DSSYK}}(\beta) = \langle 0 | e^{-\beta \hat{T}} | 0 \rangle.
\end{equation}
The spectrum and eigenstates of the transfer matrix \eqref{transfer} are known, enabling us to compute the partition function exactly, as we will shortly demonstrate in \ref{semi}.

Matter operators \(\mathcal{O}_\Delta\) can also be introduced in DSSYK \cite{Berkooz:2018jqr} and are defined as
\begin{equation}\label{matter}
    \mathcal{O}_\Delta = \mathrm{i}^{\Delta p/2} \sum_{i_1 < \dots < i_{\Delta p}} M_{i_1 \dots i_{\Delta p}} \psi_{i_1} \dots \psi_{i_{\Delta p}}\,,
\end{equation}
where \(M_{i_1 \dots i_{\Delta p}}\) are (Gaussian) random variables drawn independently from the random couplings of the Hamiltonian and the dimension \(\Delta\) of the operator is related to the number of interacting fermions in \eqref{matter}, with \(0 < \Delta < 1\).
The two-point correlation function of \eqref{matter}, with the matter operators placed at $\tau_1=\tau$ and $\tau_2=\beta-\tau$, is computed by $\langle \mathcal{O}_\Delta(\tau) \mathcal{O}_\Delta(\beta-\tau) \rangle= \text{Tr}\big(e^{-\tau H_\text{SYK}}\mathcal{O}_\Delta e^{-(\beta-\tau) H_\text{SYK}}\mathcal{O}_\Delta\big)$. In the double scaling limit, this observable can be evaluated once again using the auxiliary quantum mechanical system \eqref{transfer} as a transition matrix element: 
\begin{equation}
    \langle \mathcal{O}_\Delta(\tau) \mathcal{O}_\Delta(\beta-\tau) \rangle=\bra{0}e^{-\tau \hat{T}}\,e^{-2\Delta \abs{\log q} \hat{n}}\,e^{-(\beta-\tau)\hat{T}} \ket{0}\,.\label{1.12 twopoint}
\end{equation}
We point out that, at this stage, the transfer matrix serves merely as a combinatorial tool to solve the system in the double-scaled regime. However, we will later uncover a physical interpretation of this auxiliary quantum mechanics in terms of a dual bulk gravitational model.

\subsection{Sine-dilaton gravity and DSSYK: the duality}\label{duality}
In \cite{Blommaert:2024ydx,Blommaert:2023opb} a new holographic duality was proposed between DSSYK and sine-dilaton gravity. The latter is a dilaton gravity theory in two dimensions with a sine profile for the dilaton potential. It is described by the following path integral: 
\begin{equation}\label{b+b}
    \int \mathcal{D} g \mathcal{D}\Phi\,\exp\bigg( \frac{1}{2}\int \mathrm{d} x \sqrt{g}\bigg(\Phi R+\frac{\sin(2\abs{\log q} \Phi)}{\abs{\log q}}\bigg)+\int \mathrm{d}\tau \sqrt{h} \bigg(\Phi K-\mathrm{i} \,\frac{e^{-\mathrm{i} \abs{\log q}\Phi}}{2\abs{\log q}}\bigg)\bigg)\,,
\end{equation}
where the boundary terms correspond to the usual GHY term and the appropriate counterterm required for holographic renormalization \cite{Blommaert:2024ydx}. 

We will now briefly review how the canonical quantization of sine-dilaton gravity \eqref{b+b} exactly reproduces the auxiliary quantum mechanics \eqref{transfer} of DSSYK. In order to do that, we need to classify the classical phase space of the theory we are going to quantize.
To find the classical solutions, it is useful to rescale $ 2\abs{\log q} \Phi \rightarrow \Phi$ and minimize the rescaled action
\begin{equation}\label{sine_path}
   \frac{1}{2 \abs{\log q}}\left\{\frac12\int \mathrm{d} x \sqrt{g}\,\big(\Phi R+2\sin(\Phi)\big)+\int \mathrm{d}\tau \sqrt{h}\,\big(\Phi K-\mathrm{i} \,e^{-\mathrm{i} \Phi/2}\big)\right\}\,,
\end{equation}
that, because of the overall $1/\abs{\log q}$, is characterized by a reliable semiclassical regime  when $\abs{\log q}\ll1$. \footnote{Since $\abs{\log q}$ plays the role of $G_{N}$ in \eqref{sine_path}, the $\abs{\log q}$ expansion of DSSYK corresponds to a semiclassical gravity expansion in the gravity model \eqref{sine_path}.} The classical solutions for 
a general dilaton gravity model have been classified \cite{Gegenberg:1994pv,Witten:2020ert} in a gauge where the dilaton parametrizes the radial direction $\Phi=r$. For the sine potential, the solution for the metric takes the form
\begin{align}
        \mathrm{d} s^2&=F(r)\mathrm{d} \tau^2+\frac{1}{F(r)}\mathrm{d} r^2\,,\qquad F(r)=-2 \cos(r)+2 \cos(\theta)\,. \label{metrrr}
\end{align}
The geometry exhibits a black hole horizon at $r=\theta$ and a cosmological horizon at $r=2\pi-\theta$, corresponding to surfaces with minimal and maximal areas, respectively. Actually the metric above suggests there is an infinite set of black holes horizons at $r=\theta+2\pi n$ and an infinite set of cosmological horizons at shifted locations as well. As analyzed in \cite{Blommaert:2024ydx,Blommaert:2024whf}, the role of these infinite copies of the original geometry is crucial for recovering the full spectral density of DSSYK, as it is reproduced by an infinite set of saddles on the gravity side. However, for the purposes of this paper, we will focus only on the 'original' patch of the geometry. 
The theory \eqref{sine_path} is complemented by the following boundary conditions \cite{Blommaert:2024ydx}:
\begin{equation}
\sqrt{F}\,e^{\mathrm{i} \Phi_\text{bdy}/2}=\mathrm{i}\,,\quad\Phi_\text{bdy}=\frac{\pi}{2}+\mathrm{i} \infty\,.\label{bcrescaled}
\end{equation}
Notice the asymptotics of the metric \eqref{metrrr} is consistent with \eqref{bcrescaled} as we expand around the boundary location $\Phi_\text{bdy}$. The first condition in \eqref{bcrescaled} is a Brown-Henneaux-type boundary condition, which can be motivated in the context of the \textit{Poisson sigma model} reformulation of sine-dilaton gravity, which we review in Appendix \ref{Poisson}. The boundary location will instead by justified around equation \eqref{length}.

The ADM energy of the gravity theory can then be easily identified as the subleading part of the metric \eqref{metrrr} which remains finite as we approach the holographic boundary, yielding \footnote{We have reintroduced the dependence on $\abs{\log q}$, i.e. $F(\Phi_{\mathrm{bdy}})=-e^{-\mathrm{i}\Phi_{\mathrm{bdy}}}+4 \abs{\log q}E$. } 
\begin{equation}
     E_{\mathrm{ADM}}=-\frac{\cos(\theta)}{2\abs{\log q}}\,,
  \label{energy}
\end{equation}
once we restore the dependence on $\abs{\log q}$.
We are now in a position to characterize the classical phase space of the theory. Given the structure of the metric \eqref{metrrr}, the black hole horizon area $\theta$ naturally emerges as a phase space variable in sine-dilaton gravity. However, as in JT gravity \cite{Harlow:2018tqv}, we expect a two-dimensional phase space for this theory. To determine the second phase space variable, we observe that the Lorentzian form of two classical metrics, both described by \eqref{metrrr} and sharing the same $\theta$, can still differ due to the amount of Lorentzian time evolution $T$ of two-sided spatial slices in their Kruskal extension. \footnote{Different values of $T$ will correspond to distinct initial conditions in sine-dilaton gravity.}

The classical phase space is thus spanned by $\left(\theta,T\right)$ and is characterized by the following symplectic measure:
\begin{equation}\label{symplectic}
\omega=\mathrm{d}T \wedge \mathrm{d}H_{\mathrm{grav}}=\frac{\sin(\theta)}{2 \abs{\log q}} \ \mathrm{d}T \wedge \mathrm{d}\theta
\end{equation}
where we identified the gravitational Hamiltonian $H_{\mathrm{grav}}$ with the ADM energy \eqref{energy}.

Following \cite{Harlow:2018tqv}, we will now introduce more convenient variables in the classical phase space. Because of the form \eqref{semicl_twopoint} of the semiclassical DSSK correlator looking the same as a correlator on a AdS$_2$ background, it is natural to look for a length variable that probes the latter geometry. This is given by the Weyl-rescaled length $L$ \cite{Blommaert:2024ydx}:
\begin{equation}\label{weyl}
L=\int \mathrm{d}s e^{-\mathrm{i}\Phi/2} \,.
\end{equation}
Indeed by setting 
\begin{equation}\label{change}
    r=\frac{\pi}{2}+\mathrm{i} \log (\rho+\mathrm{i}\cos(\theta))
\end{equation}
the Weyl rescaling \eqref{weyl} takes the sine-dilaton metric \eqref{metrrr} to the effective metric
\begin{equation}\label{Ads}
    \mathrm{d} s^{2}_{\mathrm{eff}}=e^{-\mathrm{i}\Phi}\mathrm{d}s^2=\left(\rho^2-\sin(\theta)^2 \right)\mathrm{d} \tau^2+\frac{\mathrm{d}\rho^2}{\rho^2-\sin(\theta)^2}  \,,
\end{equation}
which corresponds to an AdS$_2$ background. One can thus compute the length of the two-sided ERB associated to the Weyl-rescaled AdS$_2$ black hole \eqref{Ads}, which is known to be given by:
\begin{equation}\label{length}
e^{-L}=\frac{\sin(\theta)^2}{\cosh \left(\sin(\theta) T/2\right)^2} \,.
\end{equation}
The length \eqref{length} is calculated by sending the holographic boundary to asymptotic infinity, following the standard procedure in AdS$_2$ holography. The position of the holographic screen in sine-dilaton gravity can then be inferred by taking $\rho\rightarrow +\infty$ in \eqref{change}, which justifies the boundary condition chosen in \eqref{bcrescaled}.
Interestingly, by performing the Weyl rescaling directly at the level of the action, one can rewrite the sine-dilaton gravity in a form where the classical solutions are directly \eqref{change} and \eqref{Ads}, which are a real solution for the metric and a complex solution for the dilaton field.
\begin{equation}
S = \frac{1}{\abs{\log q}} \bigg[ \int dx \sqrt{g} \big( R\Phi - i e^{i\Phi} \Phi \nabla^2 \Phi + 2 \operatorname{sin} \Phi \big) + \int dx \sqrt{h} \big( K \Phi + \frac{i}{2} e^{\frac{i\Phi}{2}} \Phi n^\rho \partial_\rho \Phi - i \big) \bigg].
\end{equation}
Now we can find the canonical conjugate to the length $L$. This can be done by inverting \eqref{length}, plugging it into the symplectic measure \eqref{symplectic} and requiring the latter to take the canonical form in terms of the new variables $L$ and $P$. One can readily show that by setting
\begin{equation}\label{momentum}
    e^{-\mathrm{i} P}=-\mathrm{i}\sin(\theta)\tanh(\sin(\theta)T/2)+\cos(\theta)\,,
\end{equation}
the symplectic form indeed becomes $\omega=\frac{1}{2\abs{\log q}}\mathrm{d} L\wedge \mathrm{d} P$.
By inverting the expressions \eqref{length} and \eqref{momentum}, one expresses the gravitational Hamiltonian $H_{\mathrm{grav}}$ in terms of $L$ and $P$, resulting in:
\begin{equation}\label{grav}
    H_\text{grav}\left(L,P\right)=-\frac{\cos(P)}{2\abs{\log q}}+\frac{1}{4\abs{\log q}}e^{iP}e^{-L}
\end{equation}
When promoting the phase space variables $L,P$ to quantum operators $\hat{L},\hat{P}$ \footnote{The symplectic measure  induces the following commutation relation:
$
\left[\hat{L},\hat{P}\right]=2\mathrm{i} \abs{\log q}
$,
which is indeed consistent with the DSSYK commutation relation $\left[\hat{n},\hat{\alpha}\right]=-\hat{a}$ \cite{Lin:2022rbf}. This can be easily proven using that $\left[A,e^{x B}\right]=x e^{x B} \left[A,B\right]$.} and comparing the gravitational Hamiltonian \eqref{grav} of sine-dilaton gravity with the double-scaled SYK transfer matrix \eqref{transfer}, we establish the following dictionary:
\begin{equation}\label{grav2}
\hat{L} \equiv 2 \abs{\log q} \, \hat{n}, \qquad \hat{\alpha} = -e^{\mathrm{i}\hat{P}} ,
\end{equation}
which shows that the two systems are governed by the same Hamiltonian, and are therefore equivalent \footnote{The semiclassical limit of the DSSYK transfer matrix, for \(\abs{\log q} \ll 1\), reduces to the classical gravitational Hamiltonian \eqref{grav} in our conventions as 
$\hat{T} \xrightarrow[\abs{\log q}\rightarrow 0]{} \frac12 H_{\mathrm{grav}}$, which explains why our definition of the inverse temperature is related by $\beta_{\mathrm{here}}=2 \beta_{\mathrm{there}}$ \cite{Goel:2023svz}. Actually, as explained in \cite{Blommaert:2024ydx}, the two hamiltonians agree up to a harmless similarity transformationi which preserves the identity. }
.

\subsection{DSSYK semiclassics and one-loop corrections}\label{semi}
In this section, we analyze the semiclassical approximation (\(\abs{\log q} \ll 1\)) and the first one-loop correction to the DSSYK partition function \eqref{part} and the matter two-point function \eqref{1.12 twopoint}. These quantities will be our primary focus for comparison with the corresponding results in the gravitational model. We will first present the exact expressions for DSSYK and then explore their behavior in the limit \(\abs{\log q} \rightarrow 0\).

The eigenvalue equation associated with the transfer matrix $\hat{T}$ \eqref{transfer} is a difference equation. The spectrum of $\hat{T}$ is continuous and bounded, and can be parameterized by an angle $\theta \in \left[0, \pi\right]$. Specifically, we have:
\begin{equation}
\hat{T} \ket{\theta} = E(\theta) \ket{\theta}, \qquad E(\theta) = \frac{2 \cos(\theta)}{\sqrt{2 \abs{\log q} (1 - q^2)}}.
\end{equation}
The eigenvectors $\ket{\theta}$ are given in the chord basis by the continuous q-Hermite polynomials, i.e. $\langle \theta | n \rangle=(1-q^2)^{n/2}H_{n}\left(\cos(\theta)|q^2\right)$\footnote{One can work with with the Hermitian version of \eqref{transfer}, as for instance is done in \cite{Lin:2022rbf}, or compute separately the left and right eigenvectors 
\begin{equation}
\langle \theta | n \rangle=(1-q^2)^{n/2}H_{n}\left(\cos(\theta)|q^2\right) \qquad \langle n | \theta \rangle=(1-q^2)^{-n/2}\frac{H_{n}\left(\cos(\theta)|q^2\right)}{\left(q^2;q^2\right)_n}
\end{equation}
as in \cite{Blommaert:2023opb}. Here $\left(a;q^2\right)_{n}=\prod_{k=0}^{n-1}\left(1-a q^{2k}\right)$ denotes the q-Pochhammer symbol.
}.
By expanding \eqref{part} on the eigenvectors basis, the DSSYK partition function is given by \cite{Berkooz:2018jqr,Berkooz:2018qkz}
\begin{equation}
\label{labelless}
    Z_\text{DSSYK}(\beta)=\int_0^\pi \mathrm{d}\theta\, (q^2,e^{\pm 2\i\theta};q^2)_\infty\,\exp\bigg(-\beta \frac{\cos(\theta)}{\sqrt{2\abs{\log q}\left(1-q^2\right)}}\bigg)\,,
\end{equation}
where the DSSYK spectral density $\rho(\theta)=(q^2, e^{\pm 2\i\theta};q^2)_\infty$ is determined by ensuring the completeness relation $\int_{0}^{\pi}\mathrm{d}\theta  \rho(\theta)\ket{\theta}\bra{\theta}=\mathbb{1}$ \footnote{We can compute the spectral density from the overlap
\begin{equation}
\langle \theta | \theta' \rangle=\sum_{n=0}^{+\infty}\langle \theta | n\rangle \langle n |\theta' \rangle=\sum_{n=0}^{+\infty} \frac{1}{\left(q^2;q^2\right)_n}H_{n}\left(\cos(\theta)|q^2\right)H_{n}\left(\cos(\theta')|q^2\right)=\frac{\delta \left(\theta-\theta'\right)}{(e^{\pm 2\i\theta};q^2)_\infty}
\end{equation}.}.
The two-point function of DSSYK matter operators $\mathcal{O}_\Delta$ can also be computed exactly, by inserting two completeness relations in \eqref{1.12 twopoint} and obtaining
\begin{equation}\label{2exact}
\begin{split}
    &\langle \mathcal{O}_\Delta(\tau) \mathcal{O}_\Delta(\beta-\tau) \rangle=\frac{1}{Z(\beta)} \times\\
    &\int_0^\pi \mathrm{d}\theta_1\,\rho(\theta_1)\int_0^\pi \mathrm{d}\theta_2\,\rho(\theta_2)\exp\bigg(-\frac{(\beta-\tau)\cos(\theta_1)}{\sqrt{2\abs{\log q}\left(1-q^2\right)}}-\frac{\tau\cos(\theta_2)}{\sqrt{2\abs{\log q}\left(1-q^2\right)}}\bigg) \bra{\theta_1} q^{2 \Delta\hat{n}} \ket{\theta_2}\,,
\end{split}
\end{equation}
where the final matrix element yields \cite{Berkooz:2018jqr,Berkooz:2018qkz,Blommaert:2023opb}:
\begin{equation}
    \bra{\theta_1} q^{2 \Delta\hat{n}} \ket{\theta_2}=\frac{(q^{4\Delta};q^2)_\infty}{(q^{2\Delta}e^{\pm\i\theta_1\pm\i\theta_2};q^2)_\infty} \,.
\end{equation}
The semiclassical analysis of the DSSYK partition function and two-point function has been covered in \cite{Goel:2023svz,Okuyama:2023bch}. Specifically, the leading order and first one-loop correction can be extracted by performing a saddle point approximation of the exact integrals \eqref{labelless}  and \eqref{2exact} in the limit $\abs{\log q}\ll 1$. Here we will not repeat the derivation but just quote the results.
The semiclassical approximation for the DSSYK partition function is:
\begin{equation}\label{semicl_part}
Z_{\mathrm{cl}}=\exp \left[-\frac{\left(\frac{\pi }{2}-\theta \right)^2-(\pi -2 \theta ) \cot (\theta )}{\abs{\log q}}\right]\,,
\end{equation}
while the one-loop correction is given by \cite{Goel:2023svz}
\begin{equation}\label{loop}
Z_{\mathrm{one-loop}}=\frac{\sin (\theta ) }{\sqrt{1+ (\frac{\pi}{2} -\theta ) \cot (\theta )}} e^{ (\frac{\pi}{2} -\theta ) \cot (\theta )}\,,
\end{equation}
The semiclassical approximation for the DSSYK two-point function is:
\begin{equation}\label{semicl_twopoint}
\langle \mathcal{O}_\Delta(\tau) \mathcal{O}_\Delta(\beta-\tau) \rangle_{\mathrm{cl}}= \frac{\sin(\theta)^{2\Delta}}{\sin(\sin(\theta)\tau/2+\theta)^{2\Delta}}\,\,.
\end{equation}
Including the one-loop correction, one obtains
\begin{equation}\label{compa}
\langle \mathcal{O}_\Delta(\tau) \mathcal{O}_\Delta(\beta-\tau) \rangle= \left.\langle \mathcal{O}_\Delta(\tau) \mathcal{O}_\Delta(\beta-\tau) \rangle\right|_{\mathrm{cl}}\bigg(1+\abs{\log q}\left(\Delta^2 \mathcal{I}+\Delta \mathcal{A}\right)+\mathcal{O}(\abs{\log q}^2)\bigg)\,.
\end{equation}
where $\mathcal{I}$ and $\mathcal{A}$ are non-trivial functions of $\tau$ and $\theta$, given by \cite{Okuyama:2023bch}:
\begin{equation}\label{one2}
\begin{split}
    \mathcal{I}&=-\frac{(\tan (\zeta )+\tan (\zeta ) (\zeta +u) \tan (u)+\tan (u)) (\tan (\zeta )+\tan (u) (\tan (\zeta ) (u-\zeta )-1))}{u \tan (u)+1} \\
    \mathcal{A}&=\frac{1}{2(1+ u \tan{u})}\left[-\frac{(1+ u \tan{u})^2}{\cos{(\zeta)}^2} + \frac{(1+ \zeta \tan{\zeta})^2}{\cos{(u)}^2} + \zeta^2(\tan{u}^2 - \tan{\zeta}^2)- \frac{1+ \zeta \tan{\zeta}}{1+ u \tan{u}}+1\right],
    \end{split}
\end{equation}
where we defined $u= \frac{\pi}{2}- \theta$ and $\zeta= \frac{\pi}{2} - \theta - \frac{\tau}{2}\sin{\theta}$. We will recover the one-loop expressions \eqref{loop} and \eqref{one2} from the gravity side in \ref{one}.
\section{Path integral formulation of sine-dilaton gravity}\label{3}
In the case of JT gravity, the gravitational path integral reduces to a description in terms of Schwarzian quantum mechanics \cite{Maldacena:2016upp,Engelsoy:2016xyb,Jensen:2016pah,Mertens:2022irh}. The path integral over the Schwarzian mode enables the calculation of several important gravitational observables at the disk level. 
In this section, we provide a path integral formulation of sine-dilaton gravity, which will serve as the framework for the computations presented in sections \ref{dete} and \ref{one}, the main results of this work. 

Analogously to the JT case, one can construct a path integral formulation for the quantum mechanics of sine-dilaton gravity, given by \eqref{grav}. Following Feynman prescription, this is given by 
\begin{equation}\label{path1}
    \mathcal{Z}_{\mathrm{grav}}^{\mathrm{naive}} = \int \mathcal{D}\varphi \mathcal{D}p \,\,\ \exp\left\{ \frac{1}{|\log q|} \int_{0}^{\beta} \mathrm{d}\tau \left( \mathrm{i} p \frac{\mathrm{d}}{\mathrm{d}\tau} \varphi + \frac{1}{2} \cos(p) - \frac{1}{4} e^{\mathrm{i} p } e^{- 2 \varphi} \right) \right\},
\end{equation}
where the action just follows from the Legendre transform of the Hamiltonian \eqref{grav} and, for convenience, we defined $L\equiv 2 \varphi$. Based on the derivation presented in \ref{duality}, equation \eqref{path1} captures the full gravitational path integral \eqref{b+b} for sine-dilaton gravity, just as the Schwarzian path integral computes the JT gravity partition function \cite{Mertens:2017mtv,Stanford:2017thb}.

Notably, the path integral \eqref{path1} was actually derived from the first time in \cite{Blommaert:2023opb} in the context of quantum groups. There, the same quantum system above is embedded into a more general theory with 6 fields $\left(\phi,\beta,\gamma,p_{\phi},p_{\beta},p_{\gamma}\right)$, which describes the motion of a particle on $SU_{q}(1,1)$ \footnote{The fields $\left(\phi,\beta,\gamma\right)$ represent the coordinates over the quantum group manifold.}.
This more general theory reduces to the so-called q-Liouville theory, which exactly corresponds to the path integral \eqref{path1}, when two constraints are imposed on the classical phase space of the group manifold \footnote{These constraints allow one to gauge-fix $\beta=\gamma=0$ and, in doing so, reduce the theory to the form \eqref{path1}.}. In the gravity context, these constraints on the quantum group implement two asymptotic Brown-Henneaux-type boundary conditions of the form \eqref{bcrescaled} \cite{Blommaert:2023opb}. We will discuss the precise correspondence of these quantities in Section \ref{Poisson}, where the link between the first order formulation of sine-dilaton gravity and the quantum group $SU_{q}(1,1)$ is made manifest.

Since sine-dilaton gravity can be viewed as a particular deformation of JT gravity \footnote{The leading term in the expansion of the dilaton potential in the action \eqref{b+b} when $\abs{\log q}\ll 1$ yields the linear dilaton potential of JT gravity. To recover JT in the quantum amplitudes, one also has to zoom in the region $\theta\ll 1$.}, the associated path integral \eqref{path1} can be interpreted as the corresponding deformation of the boundary path integral. One can, in fact, rescale $p\rightarrow p \abs{\log q}$ in the action \eqref{path1}, expand as $\abs{\log q}\ll 1$ and integrate out $p$, recovering the usual Liouville action for $\varphi$, corresponding to the two-sided formulation of JT gravity \cite{Blommaert:2023opb,Bagrets:2016cdf,Harlow:2018tqv} \footnote{In this language, the ordinary Schwarzian is the one-boundary description of JT gravity}. Unlike JT though, due to the structure of the action \eqref{path1}, it is not possible to simply integrate out the momentum and reduce the system to a single boundary mode. As we shall see, this complicates computations in this theory.

Before moving on, we should actually consider a modification of the gravitational path integral \eqref{path1} if we aim to recover the DSSYK amplitudes. A puzzling aspect of the duality between sine-dilaton gravity and DSSYK is that, despite both systems being described by the same Hamiltonian, as shown in \ref{duality}, a naive semiclassical analysis of the sine-dilaton path integral \eqref{b+b} does not align with the corresponding analysis in DSSYK. This mismatch arises from the difference between the microscopic temperature of DSSYK:
\begin{equation}\label{beta}
\beta_{\mathrm{DSSYK}} = \frac{2\pi - 4\theta}{\sin(\theta)}
\end{equation}
and the so-called "fake" temperature \cite{Lin:2023trc,Lin:2022nss}
\[
\beta_{\mathrm{fake}} = \frac{2\pi}{\sin(\theta)},
\]
which corresponds, in the gravity framework, to the Hawking temperature associated with the black hole background metric \eqref{metrrr}. This issue has been extensively addressed in \cite{Blommaert:2024ydx}. The resolution relies on a different choice of the Hartle-Hawking vacuum for the gravity theory. Naively, one might associate this vacuum with the state $\ket{-\infty}$, because the length $L$, as computed in \eqref{length}, corresponds to the holographically renormalized geodesic distance in the effective AdS$_2$ geometry \eqref{Ads}, which approaches $-\infty$ as the Euclidean separation of the boundary operators approaches $L_{\mathrm{real}} = 0$. Therefore, the naive partition function $\bra{-\infty}e^{-\beta H_{\mathrm{grav}}}\ket{-\infty}$, associated with the path integral \eqref{path1}, would reproduce the incorrect result with the fake temperature, which fails to match DSSYK.

On the other hand, the dictionary \eqref{grav2} and the DSSYK partition function \eqref{part} suggest choosing $\ket{L=0}$ as the Hartle-Hawking vacuum state. This choice represents a non-trivial geometric constraint for the gravity theory. The justification for this different choice of the Hartle-Hawking vacuum will be rigorously provided in upcoming work \cite{Blommaert:2024whf}, where gauging a global symmetry in the bulk theory \eqref{b+b} will be shown to render all negative-length states $\ket{L < 0}$ as null states, thus projecting them out of the physical Hilbert space.

While we will not delve into these details here, we state that the correct modification of the naive gravitational path integral, which matches DSSYK at temperature \eqref{beta}, is:
\begin{equation}\label{path}
    \mathcal{Z}_{\mathrm{grav}} = \int_{\varphi(0)=\varphi(\beta)=0} \mathcal{D}\varphi \mathcal{D}p \,\, \exp\left\{ \frac{1}{|\log q|} \int_{0}^{\beta} \mathrm{d}\tau \left( \mathrm{i} p \frac{\mathrm{d}}{\mathrm{d}\tau} \varphi + \frac{1}{2} \cos(p) - \frac{1}{4} e^{\mathrm{i} p} e^{-2 \varphi} \right) \right\}
\end{equation}
where we explicitly impose the boundary condition $\varphi(0) = \varphi(\beta) = 0$, which ensures the correct periodicity $\beta$ for the thermal circle.

In DSSYK, we also introduced the two-point function \eqref{1.12 twopoint} of matter operators \eqref{matter}. We can ask ourselves what the dual observable is in the gravity theory \eqref{b+b}, which turns out to be the boundary-to-boundary propagator of a massive bulk probe that couples to the length \( L \) \eqref{length}. The action of this probe is given by \cite{Blommaert:2024ydx}:
\begin{equation}\label{non_minimal}
S_{\mathrm{matter}}=\int \mathrm{d}^2 x \sqrt{g}\,\left(g^{\mu\nu}\partial_\mu \chi \partial_\nu \chi +m^2 e^{-2\mathrm{i} |\log q| \Phi}\chi^2\right)= \int \mathrm{d}^2 x \sqrt{g_\text{eff}}\,\left(g^{\mu\nu}_\text{eff}\partial_\mu \chi \partial_\nu \chi +m^2 \chi^2\right)\,,
\end{equation}
which describes a scalar field non-minimally coupled to the metric, meaning it also couples to the dilaton, as shown in \eqref{non_minimal}. This coupling ensures that the matter field probes the effective metric \eqref{Ads}, which corresponds to an AdS$_2$ geometry. According to standard AdS/CFT holography, the dual of the boundary-to-boundary propagator of the matter field \(\chi\) is represented by the bilocal operator \( e^{-2 \Delta \varphi} \) \cite{Blommaert:2018oro,Blommaert:2023opb} \footnote{The usual AdS/CFT relation between the mass and the conformal dimension $\Delta$ holds. }.
The expectation value of \( e^{S_{\mathrm{matter}}} \) in the gravitational path integral \eqref{b+b}, in the regime where the probe is light and does not backreact on the geometry, is thus computed by inserting \( e^{-2 \Delta \phi} \) into the path integral \eqref{path}, i.e.,
\begin{equation}\label{path2}
\langle e^{S_{\mathrm{matter}}}\rangle_{\mathrm{grav}} \simeq \int_{\varphi(0)=\varphi(\beta)=0} \mathcal{D}\varphi \mathcal{D}p \,\, e^{-2 \Delta \varphi} \ \exp\left\{- \frac{S_{\mathrm{qLiouville}}\left(\varphi,p\right)}{|\log q|}  \right\},
\end{equation}
where the symbol \(\simeq\) indicates that we consider only the saddle-point approximation of the matter path integral, as is typical in the AdS/CFT context.

In the following sections, we will test both \eqref{path1} and \eqref{path2} at the semiclassical and one-loop levels.

\subsection{Semiclassical comparison}
The semiclassical analysis of the path integral \eqref{path} was perfomed in \cite{Blommaert:2024ydx}, showing a perfect matching with the DSSK semiclassics \ref{semi} \footnote{A similar analysis was done in \cite{Blommaert:2023wad} in the case of Liouville gravity, where $q=e^{\pi \mathrm{i}b^2}$ and the dual boundary theory is still described by a version of the q-Schwarzian, though with different boundary conditions. }. In this section, we briefly recall the basic ingredients of the on-shell action derivation, before focusing on one-loop corrections in the following sections. The equations of motion derived from the q-Liouville action \eqref{path} are
\begin{equation}
    \frac{\mathrm{d}^2}{\mathrm{d} \tau ^2}\varphi=-\frac{1}{4}e^{-2\varphi}\,,\quad \frac{\mathrm{d}}{\mathrm{d}\tau}e^{-\mathrm{i}p}=-\frac{1}{2}e^{-2\varphi}\,.
\end{equation}
The first equation represents Liouville's equation, whose solutions correspond to AdS$_2$ conformal factors. This is consistent with interpreting $\varphi$ as the length of the Einstein-Rosen bridge (ERB) in the effective AdS$_2$ geometry \eqref{Ads}. Introducing an integration constant $\theta$, the classical solutions are given by:
\begin{align}\label{sol}
    e^{-2\varphi} = \frac{\sin(\theta)^2}{\sin(\sin(\theta)\tau/2+\theta)^2}\,, \quad e^{-\mathrm{i}p} = \frac{\sin(\theta)}{\tan(\sin(\theta)\tau/2 + \theta)} + \cos(\theta)\,.
\end{align}
Notice the solution for $\varphi$ is consistent with the boundary conditions $\varphi(0)=\varphi\left(\frac{2\pi-4\theta}{\sin (\theta)}\right)=0$, as it should.
On the solutions \eqref{sol}, one can explicitly verify that the Hamiltonian is constant:
\begin{equation}
    E_{\mathrm{grav}} = -\frac{\cos(p)}{2\abs{\log q}} + \frac{1}{4\abs{\log q}} e^{\mathrm{i} p} e^{-2\varphi} = -\frac{\cos(\theta)}{2\abs{\log q}}\,.\label{2.22 en}
\end{equation}
This indeed corresponds to the ADM energy \eqref{energy} of the gravitational model.

The non-trivial contribution to the on-shell action is the entropy part. 
Introducing the complex variable \( z = e^{\mathrm{i} \sin(\theta)\tau} \), the semiclassical entropy can be rewritten as \cite{Blommaert:2024ydx}:
\begin{equation}\label{g(z)}
S_{\mathrm{grav}}=\frac{1}{\abs{\log q}} \int_{0}^{\beta} \mathrm{d}\tau \left( \mathrm{i} p \varphi'\right)=\frac{1}{2|\log q|} \int_{\gamma} \mathrm{d}z \, \left( \frac{1}{z} + \frac{2 e^{2 \mathrm{i} \theta}}{1 - e^{2 \mathrm{i} \theta} z} \right) \log \left( \frac{e^{-\mathrm{i} \theta} - e^{3 \mathrm{i} \theta} z}{1 - e^{2 \mathrm{i} \theta} z} \right),
\end{equation}
where \(\gamma\) is a contour starting from \( z=1 \) and ending at \( z = e^{-4\mathrm{i}\theta} \), encircling the pole at \( z = 0 \).
The integral can be explicitly evaluated and yields \footnote{In the last step we made use of the identity $
    \text{Li}_2(x)+\text{Li}_2(1/x)=-\frac{\pi^2}{6}-\frac{1}{2}\log^2(-x)\,,\label{Li2}
$}
\begin{equation}\label{I4}
\begin{split}
\abs{\log q}S_{\mathrm{grav}} &= \frac{\pi^2}{12} + \frac{1}{2} \mathrm{Li}_{2}\left(e^{-2\mathrm{i}\theta}\right) + \frac{1}{2} \mathrm{Li}_{2}\left(e^{2\mathrm{i}\theta}\right)  + \frac12 \log(-e^{2 \mathrm{i}\theta})^2 \\
&=\frac{1}{4}\log(-e^{2 \mathrm{i} \theta})^2=-\left(\frac{\pi}{2}-\theta\right)^2\,.
\end{split}
\end{equation}
Combining the energy and entropy terms \eqref{2.22 en} and \eqref{I4}, we conclude the semiclassical free energy for sine-dilaton gravity is given by:
\begin{equation}
\mathcal{F}_{\mathrm{cl}}=-\log (Z_{\mathrm{cl}})=-S_{grav}+\beta H_{\mathrm{grav}}=\frac{\left(\frac{\pi}{2}-\theta\right)^2}{\abs{\log q}}-\frac{}{}\frac{(\pi-2\theta)\cot(\theta)}{\abs{\log q}}
\end{equation}
This indeed matches with the semi-classical partition function of DSSYK \eqref{semicl_part}.

Moreover, the semiclassical approximation of the path integral \eqref{path2} amounts to evaluate the bilocal operator $e^{-2\Delta \varphi}$ on the classical solution \eqref{sol} for $\varphi$, which yields 
\begin{equation}\label{sese}
\begin{split}
\langle e^{-2\Delta \varphi} \rangle _{\beta}&\simeq_{\abs{\log q}^0} \ e^{-2\Delta \varphi_{\mathrm{cl}}(\tau)}=\left(\frac{\sin(\theta)^2}{\sin \left(\sin(\theta)\tau/2+\theta\right)^2}\right)^{\Delta} 
\end{split}
\end{equation}
which shows perfect agreement with the semiclassical DSSYK matter correlator \eqref{semicl_twopoint}.
\subsection{The one-loop determinant for the partition function}\label{dete}
In the previous section we reviewed how the duality works at the semiclassical level.
Now, we aim to take a step further and evaluate the first quantum correction to the q-Liouville path integral. To achieve this, we expand around the classical solutions \eqref{sol}:
\begin{equation}\label{quantum}
		\Phi_i=\Phi_{i}^{(\mathrm{cl})}+\sqrt{\abs{\log q}} \: \delta \Phi_{i} 
\end{equation}
where we defined: 
\begin{equation}
	\Phi_1 := \varphi \quad ; \quad \Phi_2 := p \quad ; \quad \delta\Phi_1 := \delta\varphi \quad ; \quad \delta\Phi_2 := \delta p
\end{equation}
and we regard $\delta \Phi_{i}$ as the quantum fluctuations.
As a result, the action has the following functional expansion:
\begin{equation}
	\frac{\mathcal{S}}{\abs{\log q}}=\frac{\mathcal{S}_{\mathrm{cl}}}{\abs{\log q}} + \frac{1}{\sqrt{\abs{\log q}}}\left.\int \mathrm{d} \tau \frac{\delta \mathcal{L} }{\delta \Phi_j}\right|_{\mathrm{cl}} \delta \Phi_j (\tau)+ \frac{1}{2} \left.\int \int \mathrm{d} \tau  \mathrm{d} \tau^{\prime} \frac{\delta^2 \mathcal{L}}{\delta \Phi_i \delta \Phi_j}\right|_{\mathrm{cl}} \delta \Phi_i (\tau) \: \delta \Phi_j (\tau^{\prime})+\ldots
\end{equation}
The linear term vanishes due to the equations of motion, while the second order contribution reads:
\begin{equation}
	\begin{split}
		\mathcal{S}^{(2)}=  \int\int \mathrm{d} \tau  \mathrm{d} \tau^{\prime } \: &\Bigg[ \frac{\mathrm{i}}{2}\delta p \left(\tau^{\prime }\right)  \: \delta \dot{\varphi}\left(\tau\right) \: - \frac{\mathrm{i}}{2}\:\delta \varphi \left(\tau^{\prime }\right)   \: \delta  \dot{p}\left(\tau\right)-\frac{1}{4} \cos \left(p_{\mathrm{cl}}\right) \delta p\left(\tau^{\prime }\right)  \delta p\left(\tau\right)+ \\
		&+\frac{1}{8}e^{-2 \varphi_{\mathrm{cl}}+\mathrm{i} p_{\mathrm{cl}}}\left(\delta p \left(\tau^{\prime }\right)+2 \mathrm{i} \delta \varphi\left(\tau^{\prime }\right)\right) \left(\delta \mathrm{p}\left(\tau\right)+2 \mathrm{i} \delta \varphi\left(\tau\right)\right)\Bigg]\delta(\tau-\tau')
	\end{split}
\end{equation}
The previous term is quadratic in the perturbations, as expected, and can be written as: 
\begin{equation}
	\mathcal{S}^{(2)} =  \int\int \mathrm{d} \tau \mathrm{d} \tau^{\prime} \: \delta \Phi_i (\tau^{\prime}) \:K_{ij}(\tau,\tau') \:\delta \Phi_j (\tau)
\end{equation}
where we introduced $K_{ij}(\tau,\tau')=\delta(\tau-\tau')L_{ij}(\tau)$, with $L_{ij}(\tau)$ being a matrix operator of the form:
\begin{equation}\label{L_operator}
	L_{ij} (\tau) =  \begin{pmatrix}
	 -\frac{1}{2} e^{-2 \varphi_{\mathrm{cl}}(\tau)+\mathrm{i} p_{\mathrm{cl}}(\tau)} && \frac{\mathrm{i}}{4}  e^{-2 \varphi_{cl}(\tau)+\mathrm{i} p_{\mathrm{cl}}(\tau)} - \frac{\mathrm{i}}{2}\frac{\mathrm{d}}{\mathrm{d}\tau} \\ \\
		\frac{\mathrm{i}}{4}  e^{-2 \varphi_{\mathrm{cl}}(\tau)+\mathrm{i} p_{\mathrm{cl}}(\tau)} + \frac{\mathrm{i}}{2}\frac{\mathrm{d}}{\mathrm{d}\tau} &&	-\frac{1}{4} \cos (p_{\mathrm{cl}}(\tau))+\frac{1}{8} e^{-2 \varphi_{\mathrm{cl}}(\tau)+\mathrm{i} p_{\mathrm{cl}}(\tau)}
	\end{pmatrix}.
\end{equation}
As a result, the path integral over field perturbations is gaussian and the computation of the one-loop partition function is reduced to evaluating the functional determinant of $L$ on the circle, i.e.
\begin{equation}
	\mathcal{Z}_{\mathrm{grav}}^{\mathrm{one-loop}}=e^{-\frac{\mathcal{S}\left(\Phi^{(\mathrm{cl})}\right)}{\abs{\log q}}}\int_{\delta \varphi(0)=\delta\varphi(\beta)=0} \mathcal{D}\delta \varphi 
 \ \mathcal{D}\delta p  \ e^{-\mathcal{S}^{(2)}}=\frac{ e^{-\frac{\mathcal{S}_{\mathrm{cl}}}{\abs{\log q}}}}{\sqrt{\text{Det} \left( L \right)}}.
\end{equation}
where the boundary conditions for the quantum fluctuations $\delta \varphi(0)=\delta\varphi(\beta)=0$ follow from \eqref{path1}.
Therefore, we will now focus on the computation of the aforementioned object $\det (L)$ . Naively, one approach is to extract the spectrum of the $L$ operator and compute its determinant using zeta-function regularization \cite{Hawking:1976ja, ray1971r, voros1987spectral, Elizalde:1995hck, ElizaldeOdintsov, Elizalde:1993ue, McKeon:1986rc, Dunne:2007rt}. However, due to the complexity of the operator, this procedure is not straightforward, as it is not simple to solve the spectral problem.
Since the problem is one-dimensional, we can take advantage of results along the lines of Gel'fand Yaglom's theorem \cite{gel1960integration}. The benefit of this approach is that it is not required to compute the set of eigenvalues, indeed the determinant can be written in terms of the solutions of an initial value problem.  Strictly speaking, Gel'fand Yaglom's theorem applies to Schroedinger operators on the line, with Dirichlet boundary conditions. However, the result has been extended to matrix elliptic operators of any order, with arbitrary boundary conditions \cite{Forman1987,Forman1992, McKane:1995vp,Kirsten:2003py, Kirsten:2004qv, Dunne:2007rt, Forini:2015bgo, burghelea1991determination, Braverman:2013ozc, Falco:2017ceh}. We will rely primarily on Forman's results \cite{Forman1987}, which we briefly resume below.
\paragraph{Determinant of elliptic differential operators}
Let's consider an elliptic differential operator, of the form: 
\begin{equation}\label{FormanOperator}
	\mathcal{O}=P_0(\tau) \frac{\mathrm{d}}{\mathrm{d} \tau}+P(\tau)
\end{equation}
where the $P_0 (\tau)$ and $P(\tau)$ are complex $n \cross n$ matrices on the interval $I = [a,b]$ \footnote{The operator is elliptic if $\det (P_0 (\tau)) \neq 0$ $\forall \: \tau \in I$.}. Let's admit that we are in a position to calculate the fundamental solution $Y (\tau)$ of the following homogeneous problem: 
\begin{equation}\label{Def_Fundamental_solution}
	\mathcal{O} \:Y(\tau) = 0 \quad ; \quad Y(a) = \mathds{1}
\end{equation}
where $Y(\tau)$ is an $N \cross N$ matrix constructed by stacking independent solutions (tuples) of \eqref{FormanOperator}, which satisfy the condition of having no divergence within the interval $I$.
This object provides a basis of solutions of the homogeneous problem, indeed, it is clear that if we define: 
\begin{equation}
	f(\tau) := Y(\tau ) f_a \quad ; \quad f_a \in \mathbb{C}^{N} 
\end{equation}
Then $f(\tau)$ is solution of the problem with arbitrary Cauchy boundary conditions:
\begin{equation}
	\mathcal{O} f(\tau) = 0 \quad ; \quad f(a) = f_a 
\end{equation}
Moreover, if we want to impose a more general condition on both the extremes, it is sufficient to introduce the matrices $M$ and $N$, such that: 
\begin{equation}
	M f(a) + N f(b) = 0 
\end{equation}
As in Gel'fand Yaglom's theorem, the functional determinant, up to regularization, is encoded in the chosen boundary value problem $\{\mathcal{O}, M, N\}$, through the formula \cite{Forman1987}:
\begin{equation}\label{Forman}
	\text{Det}^{(F)}(\mathcal{O}) = \left[ \exp \left(\int_{a}^{b} \text{Tr} \: [ R_{C_{\theta_1 \theta_2}}(\tau) \: P(\tau) \: P_0(\tau)^{-1} ]\:\mathrm{d}\tau \right) \det(M + N Y_{L}(b))\right]
\end{equation}
where $R_{C_{\theta_1 \theta_2}}$ is a projector defined as follows. If we consider the eigenvalues of the matrix $-i P_0$, they can belong either  to the cone $C_{\theta_1 \theta_2} = \{z \in \mathbb{C}| \: \theta_1 < z < \theta_2\}$ or its opposite $\overline {C}_{\theta_1 \theta_2}$ (for some choice of $\theta_1$ and $\theta_2$). As a result, we can define two sets of eigenvalues $\mathcal{S_C}$ and $\mathcal{S_{\overline C}}$, depending in which region they fall. Then $R_{C_{\theta_1 \theta_2}}(\tau)$ is the projector onto the subspace generated by the eigenvectors corresponding to the eigenvalues in  $\mathcal{S_C}$. The choice of angles $\theta_1$ and $\theta_2$ might seem arbitrary, however, since the number of eigenvalues is finite, there is a limited set of possible projectors R. This arbitrariness of $R$ is a remnant of the possible choices of principal angles with which to carry out analytic continuation in $\zeta$-function regularization. In our analysis it will be clear what is the proper choice of $R$.

Actually \eqref{Forman} is only formal, since it could diverge. In order to make sense of it, the strategy is to introduce a regularization scheme. Let's consider a simpler operator $\hat{\mathcal{O}}$ which shares the same principal symbol $P_0$ with $\mathcal{O}$ and for which the calculation of the spectrum and $\zeta$-function regularization is easy. Then, the $\zeta$-regularized result is given by \cite{Forman1987, Dunne:2007rt} \footnote{Alternatively, if one focuses on operators on the circle, one can rely on the results of \cite{burghelea1991determination}, who reports a formula similar to Forman's, but already regularized.}: 
\begin{equation}\label{Zeta_Regularization_Determinat}
	\text{Det}^{(\zeta)}(\mathcal{O}) = \frac{\text{Det}^{(F)}(\mathcal{O}) }{\text{Det}^{(F)}(\hat{\mathcal{O}}) } \: \text{Det}^{(\zeta)}(\hat{\mathcal{O}})
\end{equation}

\paragraph{Computation of the one-loop determinant}
We now use this machinery to analyze the problem of our interest, which is defined on the circle $I = [0, \beta]$.
By comparing the form of $L$ in \eqref{L_operator} with \eqref{FormanOperator}, we infer that $P_0$ and $P(\tau)$ are the following $2 \times 2$ matrices:
\begin{equation} \label{PPPP}
\begin{split}
		&P_0  = \left(
		\begin{array}{cc}
			0 & -\frac{i}{2} \\ \\
			\frac{i}{2} & 0 \\
		\end{array}
		\right) \qquad \qquad  P(\tau) = \left(
			\begin{array}{cc}
				  h ( \tau)& -\frac{i}{2}  h ( \tau) \\ \\ -\frac{i}{2}  h ( \tau) & -\frac{\cos (\theta )}{4}\\
			\end{array} 
			\right)   
\end{split}\end{equation}
where  
\begin{equation} 
h ( \tau) := \frac{\sin ^2(\theta )}{\cos (\tau  \sin (\theta )+3 \theta )-\cos (\theta )}
\end{equation}
is obtained  by evaluating \eqref{L_operator} on the classical solutions \eqref{sol}. At this stage, we observe that while \( h(0) \) is well-defined, \( h(\beta) \) becomes divergent. Consequently, we expect the solutions \( Y(\tau) \) to exhibit some divergence as \(\tau\) approaches \(\beta\). To address this issue, we introduce a cutoff at the upper limit of the interval, redefining it as \( I^{(\epsilon)} = [0, \beta_{\epsilon}] \). After completing the calculation, we will take the limit \( \epsilon \to 0 \) and recover the desired result, assuming no pathologies arise during this process.

The appropriate boundary conditions are implemented via the formalism: $M_{ij} \: \delta \Phi_{j}(0) + N_{ij} \:\delta \Phi_{j}(\beta_{\epsilon}) =0 $, hence, the correct choice is:
\begin{equation}\label{MN}
	M = \begin{pmatrix}
		1 && 0 \\
		0 && 1
	\end{pmatrix} , \qquad N= \begin{pmatrix}
		0 && 0 \\
		0 && -1
	\end{pmatrix}.
\end{equation}
We will face two main tasks: the calculation of the $2 \cross 2$ determinant involving the fundamental solution $Y(\beta_{\epsilon})$ and the  computation of the exponential factor in \eqref{Forman}.\\
As a first step, we need to solve the differential system $L_{ij} \: \delta \Phi_{j}= 0$. Plugging the second row of the system into the first one, we obtain the following second order differential equation for $\delta \varphi$:
\begin{equation}\label{Diff_Equation_Delta_Phi}
	h( \tau) \: \delta \varphi(\tau)+\sec(\theta)\left[-\delta \varphi^{\prime \prime}(\tau)+\delta \varphi(\tau) \:\left(h( \tau)^2+h^{\prime}( \tau)\right)\right] = 0
\end{equation}
whose general solution is:
\begin{equation}\label{Solution_delta_phi}
\delta \phi(\tau) = 	\frac{-c_1 \tau  \cos (\theta )+c_1 \csc (\theta ) \sin (\tau  \sin (\theta )+3 \theta )+c_2}{\cos (\tau  \sin (\theta )+3 \theta )-\cos (\theta )}
\end{equation}
where $c_1$ and $c_2$ are integration constants. Depending on their values, we obtain a solution of the system $ \vec{\delta \Phi}^{(c_1, c_2)} (\tau)$.
To construct the fundamental matrix $Y(\tau)$, we consider two independent solutions $ \vec{\delta \Phi}^{(1, 0)} (\tau) \: , \: \vec{\delta \Phi}^{(0,1)}(\tau)$, by setting respectively: $(c_1, c_2) = \{(1, 0) \: ; \: (0,1) \}$, and build up the invertible matrix $H(\tau)$ by stacking the two independent solution vectors as columns:  
\begin{equation}\label{HHH}
H(\tau) := (  \vec{\delta \Phi}^{(1, 0)}(\tau) \: , \: \vec{\delta \Phi}^{(0,1)}(\tau))
\end{equation}
We finally normalize it, by defining 
$Y(\tau):=H(\tau)H^{-1}(0)$,
which clearly satisfies $Y(0) = \mathds{1}$, as required in \eqref{Def_Fundamental_solution}.
We are now in position to evaluate the determinant in \eqref{Forman}:
\begin{equation}\begin{split}
	&\det (M + N Y_{L}(b))=  \\& = \frac{\csc (\theta ) \sin \left(\frac{b}{2}  \sin (\theta )\right) \left[b \cos \left(\frac{b}{2}  \sin (\theta )+3 \theta \right)-2 \csc (\theta ) \sec (\theta ) \sin \left(\frac{b}{2}  \sin (\theta )\right)\right]}{2 [\cos (\theta )-\cos (b \sin (\theta )+3 \theta )]}
	\end{split}
\end{equation}
As expected, evaluating the determinant at \(b = \beta\) results in a divergence that, however, can be regularized by introducing a small cutoff \(\epsilon\). After evaluating the determinant at the regularized endpoint \(\beta_{\epsilon}\), and retaining only the leading term in the expansion as \(\epsilon \to 0\), we obtain:
\begin{equation}\label{determinan_Forman}
    \text{det}(M + N Y_{L}(\beta_{\epsilon})) \sim -\frac{2 \cot (\theta ) ((\pi -2 \theta ) \cot (\theta )+2)}{\epsilon }.
\end{equation}
In order for the divergence to disappear it is necessary for the exponential in \eqref{Forman}, to contribute at $\epsilon$ order, this condition guides us to the choice of the correct projector $R_{C_{\theta_1 \theta_2}}(\tau)$. The eigenvalues of $- i P_0$ are $\{\frac{i}{2}, -\frac{i}{2} \}$, hence, depending on the choice of $\theta_1$ and $\theta_2$ there are four possibilities: $R \in \{\pi_{i/2}, \pi_{-i/2} , \mathds{1}, 0 \}$. In retrospect, the only significant one is:
\begin{equation}
	R (\tau) = \pi_{i/2}= \left(
	\begin{array}{cc}
		\frac{1}{2} & -\frac{i}{2} \\ \\
		\frac{i}{2} & \frac{1}{2} \\
	\end{array}
	\right)
\end{equation}
Indeed the integral that appears in \eqref{Forman} has the following form:
\begin{equation}\begin{split}
    \mathcal{I} := &\int_0^{\beta_{\epsilon}} \left(\frac{\sin ^2(\theta )}{\cos (\tau  \sin (\theta )+3 \theta )-\cos (\theta )}-\frac{\cos (\theta )}{4}\right) \, d\tau 
\end{split}\end{equation}
Performing the integral and expanding up to first order in $\epsilon$ we arrive at:
\begin{equation}\label{Exponential_Forman}
    e^{\mathcal{I}} \sim \frac{\epsilon}{4}  \:  e^{-\frac{1}{2} (\pi -2 \theta ) \cot (\theta )} \csc (\theta ) \sec (\theta )
\end{equation}
Multiplying the two results \eqref{Exponential_Forman} and \eqref{determinan_Forman}, we note that the divergence disappears and it is safe to send $\epsilon \rightarrow 0$. Since the result is already non-divergent, the only factor obtained from the strict application of \eqref{Zeta_Regularization_Determinat} is a numerical constant, which turns out to be $-1$. Hence we find:
\begin{equation}
    \text{Det}^{(\zeta)} (L)=-\text{Det}^{(F)} (L) = \frac{1}{2} e^{-\frac{1}{2} (\pi -2 \theta ) \cot (\theta )} ((\pi -2 \theta ) \cot (\theta )+2) \csc ^2(\theta ) 
\end{equation}
Including the one-loop correction, the free energy of sine-dilaton gravity is then given by 
\begin{equation}\label{loop2}
\begin{split}
\mathcal{F}_{\mathrm{one-loop}}=&\frac{\left(\frac{\pi}{2}-\theta\right)^2}{\abs{\log q}}-\frac{(\pi-2\theta)\cot(\theta)}{\abs{\log q}}+\frac12 \log \left(1+\left(\frac{\pi}{2} - \theta \right) \cot (\theta )\right)\\
&-\log (\sin (\theta))-\frac12 \left(\frac{\pi}{2}-\theta\right)\cot (\theta)
\end{split}
\end{equation}
We find a perfect match with the one-loop partition function of DSSYK in \eqref{loop}, except for an additional factor of $1/2$ for the last term in \eqref{loop2}. We point out that this last term comes from the first quantum correction to the classical ADM energy \eqref{energy} of sine-dilaton gravity and should therefore be related to different orderings for the gravitational Hamiltonian \eqref{grav}. A more refined analysis of the discretization leading to the path integral \eqref{path1} could indeed pinpoint the correct choice of ordering for the Hamiltonian \eqref{grav} leading to this small discrepancy, but we haven't investigated this aspect further in this work.

  

\section{One-loop correction to the matter two-point function}\label{one}
The aim of this section is to analyze the one-loop gravitational correction to the boundary to boundary propagator of a scalar matter field \eqref{matter}, coupled non-minimally to sine-dilaton gravity. By virtue of \eqref{path2}, this is equivalent to extracting the $\abs{\log q}$ correction to the q-Liouville path integral $\langle e^{-2\Delta \phi(\tau)} \rangle _{\beta}$, in a regime where this insertion does not back react on the geometry. After expanding each field as in \eqref{quantum}, we obtain:
\begin{equation}\label{expr}
\begin{split}
\langle e^{-2\Delta \varphi} \rangle _{\beta}&=\frac{1}{\mathcal{Z}(\beta)} e^{- \frac{\mathcal{S}\left(\Phi^{(\mathrm{cl})}\right)}{\abs{\log q}}} e^{-2\Delta \varphi_{\mathrm{cl}}(\tau)}  \ \times \\
&\int_{\delta \varphi(0)=\delta \varphi(\beta)=0} \mathcal{D} \delta \phi \mathcal{D} \delta p \ e^{-2\Delta \sqrt{\abs{\log q}}\delta \varphi(\tau)} \ \ e^{-\mathcal{S}^{(2)}-\sqrt{\abs{\log q}}\mathcal{S}^{(3)}-\abs{\log q}\mathcal{S}^{(4)}+\cdots} \ ,
\end{split}
\end{equation}
where we denoted with $S^{(n)}$ the term in the expansion of the action that contains $n$ fields. For future reference, we report below the explicit expression of $\mathcal{S}^{(3)}$:
\begin{equation}\label{S3}
\begin{split}
    \mathcal{S}^{(3)} = \int \mathrm{d} \tau &\bigg[\frac{1}{3} e^{\mathrm{i} p_{\mathrm{cl}}-2 \varphi_{\mathrm{cl}} } \delta \varphi(\tau)^3 -\frac{\mathrm{i}}{2}   e^{\mathrm{i} p_{\mathrm{cl}}-2 \varphi_{\mathrm{cl}}}\delta \varphi(\tau)^2 \delta p(\tau)-\frac{1}{4}   e^{\mathrm{i} p_{\mathrm{cl}}-2 \varphi_{\mathrm{cl}}}  \delta \varphi(\tau) \delta p(\tau)^2 \\
    & +\left(\frac{\mathrm{i}}{24}   e^{\mathrm{i} p_{\mathrm{cl}}-2 \varphi_{\mathrm{cl}} }+\frac{1}{12} \sin (p_{\mathrm{cl}})\right)\delta p(\tau)^3 \bigg]
\end{split}\end{equation}
As an initial observation regarding \eqref{expr}, we note that by neglecting all quantum corrections in \(\abs{\log q}\) beyond \(\abs{\log q}^0\), the path integral contributes only a factor of the form \(\det L^{-\frac{1}{2}}\). Dividing by the partition function at the same order, we recover the semiclassical result \eqref{sese}, as expected.
More interestingly, we aim to extend the expansion in \(\abs{\log q}\) and compute the first correction to the above result, incorporating non-trivial interacting terms from the action. At this order, the normalized contribution from the path integral takes the form:

\begin{equation}
\begin{split}
&\sim \left( \det L \right)^{\frac{1}{2}} \int \mathcal{D} \delta \varphi \mathcal{D} \delta p \, e^{-\mathcal{S}^{(2)}} \left[ 1 - 2 \Delta \delta \varphi \sqrt{\abs{\log q}} + \abs{\log q} \left( 2 \Delta^2 \delta \varphi^2 - 2 \Delta \delta \varphi \mathcal{S}^{(3)} \right) + \cdots \right],
\end{split}
\end{equation}
Notably, since the term \(\propto \abs{\log q}^{\frac{1}{2}}\) vanishes in the quadratic path integral, we find at order \(\abs{\log q}\) the following expression for the correlator:
\begin{equation}\label{3.12}
\begin{split}
\langle e^{-2\Delta \varphi} \rangle_{\beta} &\simeq e^{-2\Delta \varphi_{\text{cl}}(\tau)} \left[ 1 + 2 \abs{\log q} \left( \Delta^2 \langle \delta \varphi(\tau)^2 \rangle - \Delta \langle \delta \varphi(\tau) \mathcal{S}^{(3)} \rangle \right) +\mathcal{O}\left(\abs{\log q}^2\right)\right],
\end{split}
\end{equation}
where the expectation values are taken with respect to the quadratic action and are normalized by the one-loop determinant that contributes at order \(\abs{\log q}^{0}\) to the partition function.
In order to compute these quantities, it is convenient to analyse the following generating functional obtained by adding a source to the gaussian path integral:
\begin{equation}
\begin{split}
Z[\{J_i\}; \beta]=\int_{\delta \varphi(0)=\delta \varphi(\beta)=0} \mathcal{D} \delta \varphi \mathcal{D} \delta p \ e^{-\mathcal{S}_{(2)}+\int_{0}^{\beta} \mathrm{d}t J_i(t)\delta \Phi_i(t) },
\end{split}
\end{equation}
where at the exponent we have the following quantity:
\begin{equation}
\begin{split}
&\frac12 \int \int \mathrm{d}t \mathrm{d}t' \delta \Phi_{i}(t) K_{ij}(t,t')\delta \Phi_{j}(t')+\int \int \mathrm{d}t \mathrm{d}t' J_i(t)\delta(t-t')\delta \Phi_{i}(t')
\end{split}
\end{equation}
with $K_{ij}\left(t,t'\right)=L_{ij}(t) \delta (t-t')$ already introduced in \eqref{L_operator}.
If we now introduce the Green function $G_{ij}(t,t')$ for the quadratic operator $K_{ij}\left(t,t'\right)$, defined as:
\begin{equation}\label{verde}
\begin{split}
\int_{0}^{\beta} \mathrm{d}t'' K_{ij}\left(t,t''\right) G_{jk}(t'',t')=\delta (t-t')\delta_{ik},
\end{split}
\end{equation}
and we perform the standard change of variable inside the path integral 
\begin{equation}
\begin{split}
\delta \tilde{\Phi}_{i}(t)=\delta \Phi_i (t)+\int \mathrm{d}t' G_{ij}(t,t')J_{j}(t'),
\end{split}
\end{equation}
we immediately obtain the following result for the generating function:
\begin{equation}\label{generating}
\begin{split}
Z[\{J_i\}; \beta]=\left(\det L\right)^{-\frac12} \exp \left[- \frac12 \int_{0}^{\beta} \int_{0}^{\beta} \mathrm{d}t \mathrm{d}t' J_{i}(t)G_{ij}(t,t')J_{j}(t')\right].
\end{split}
\end{equation}
Interestingly, we observe that the determinant appearing as a prefactor to the exponential precisely cancels the normalization factor of the expectation values in \eqref{3.12}. As a result, we can disregard this determinant and treat the expectation values in \eqref{3.12} as arising from the functional differentiation with respect to the sources in the generating functional \eqref{generating}.
Using this method, it is immediate to verify that:
\begin{equation}\label{Gphiphi}
    \langle \delta \varphi(\tau)^2 \rangle = \frac{\delta}{\delta J_{\varphi}(\tau)}\frac{\delta}{\delta J_{\varphi}(\tau)} \exp \left[- \frac12 \int_{0}^{\beta} \int_{0}^{\beta} \mathrm{d}t \mathrm{d}t' J_{i}(t)G_{ij}(t,t')J_{j}(t')\right] = - G_{\varphi \varphi}(\tau, \tau).
\end{equation}
The challenging task, of course, is to compute the Green's function \( G_{ij} \) for the matrix operator \( L \) in \eqref{PPPP}, subject to the boundary conditions \(\delta \varphi(0) = \delta \varphi(\beta) = 0\) and periodic boundary conditions for \(\delta p\). However, it is possible to construct the Green's function once the solutions to the homogeneous problem are known. As derived in Appendix \eqref{general}, the Green's function can be expressed as:
\begin{equation}\label{ggg}
G(t, \tau) = \theta(t - \tau) G^{(a)}(t, \tau) + \theta(\tau - t) G^{(r)}(t, \tau),
\end{equation}

where \( G^{(a)}(t, \tau) \) and \( G^{(r)}(t, \tau) \) represent the retarded and advanced Green's functions, respectively. These are given by \footnote{Once again, we regulate the divergence of $H(\beta)$, which is exactly canceled by   $\mathcal{R}^{-1} $, by introducing the cutoff $\epsilon$ and finally sending $\epsilon \rightarrow 0$.}:
\begin{equation}\label{ar}
\begin{split}
G^{(a)}(t, \tau) &= -H(t) \left[   \mathcal{R}^{-1}  N H\left(\beta_{\epsilon}\right) - 1 \right] H^{-1}(\tau)P_{0}^{-1} \\
G^{(r)}(t, \tau) &= -H(t)  \mathcal{R}^{-1}  N H\left(\beta_{\epsilon}\right) H^{-1}(\tau)P_{0}^{-1}
\end{split}
\end{equation}
where \( H(\tau) \) is the fundamental matrix solution of the homogeneous problem \eqref{HHH},  $ \mathcal{R}$  is defined as:
\[
  \mathcal{R} = M H(0) + N H(\beta_{\epsilon}),
\]
with \( M \) and \( N \) being the matrices already introduced in \eqref{MN} to implement the boundary conditions. Using \eqref{ar}, the complete form of the Green's function for $L$ is derived in detail in \eqref{complete_form1} and \eqref{complete_form2}, but it is rather intricate. For our purposes, we just report here the $G_{\varphi \varphi}(\tau,\tau)$ component \footnote{In computing the Green's function at coincident points from \eqref{ggg}, we are assuming $\theta(0)=\frac12$.}, which is the one of our interest because of \eqref{Gphiphi}:
\begin{equation}\label{GG}
    G_{\varphi \varphi}\left(\zeta,\zeta\right)=\frac{(\tan (\zeta )+\tan (\zeta ) (\zeta +u) \tan (u)+\tan (u)) (\tan (\zeta )+\tan (u) (\tan (\zeta ) (u-\zeta )-1))}{u \tan (u)+1}
\end{equation}
where we have introduced $\zeta=\frac{\pi}{2}-\theta-\frac{\tau}{2} \sin (\theta)$ and $u=\frac{\pi}{2}-\theta$ as in \ref{semi}. We can already appreciate the matching between this quantity and $\mathcal{I}$ in \eqref{one2}.

The remaining term in \eqref{3.12} that we need to compute is \(\langle \delta \varphi(\tau) \mathcal{S}_{(3)} \rangle\). To derive an expression for this expectation value, we could, in principle, examine each term in \eqref{S3} and evaluate the corresponding integrals by systematically applying Wick contractions, as prescribed by Wick's theorem. While this method is perfectly valid, it would involve extensive calculations. Instead, we will adopt a more efficient approach that avoids these lengthy computations.
The key insight here is that it is advantageous to express \(\langle \delta \varphi(\tau) \mathcal{S}_{(3)} \rangle\) using a covariant notation and investigate the more general correlator:
\begin{equation}\label{wickwick}
\mathcal{A}_l(\tau) = \left\langle \delta \Phi_l(\tau) \int_0^{\beta} \mathrm{d}t\, T_{ijk}(t) \delta \Phi_i(t) \delta \Phi_j(t) \delta \Phi_k(t) \right\rangle,
\end{equation}
where \( T_{ijk} \) are the vertices of the cubic interaction term from the action \eqref{S3}. The non-vanishing components of the fully symmetric tensor \( T_{ijk}(t) \) can be extracted from \eqref{S3} by substituting the classical solution \eqref{sol}, which gives the following structure for the vertices:
\begin{equation}\label{T}
\begin{split}
    &T_{p p p}(t) = \frac{\mathrm{i}}{2} \cot{\left( \theta + \frac{t}{2} \sin{\theta} \right)}, \qquad T_{pp\varphi}(t)=\frac{\sin^2{(\theta)}}{6 \cos\left(3 \theta + t \sin{\theta} \right)- 6 \cos{(\theta)}},\\
    &T_{p \varphi \varphi}(t)= 2 \mathrm{i}\, T_{pp\varphi}(t), \qquad T_{\varphi \varphi \varphi}(t)= 2 \mathrm{i}\, T_{p \varphi \varphi}(t).
\end{split}\end{equation}
In the case where \( l = \varphi \), the newly defined correlator becomes \(\mathcal{A}_{\varphi}(\tau) = \langle \delta \varphi(\tau) \mathcal{S}^{(3)} \rangle\), which is precisely the quantity we wish to analyze. Similarly, for \( l = p \), we have \(\mathcal{A}_{p}(\tau) = \langle \delta p(\tau) \mathcal{S}^{(3)} \rangle\). The benefit of introducing this unified notation, treating both correlators on an equal footing, is that it enables us to derive a differential equation governing the correlator of interest, allowing us to determine its functional form.
To proceed we first observe that, by applying Wick's theorem to \eqref{wickwick}, we obtain:
\begin{equation}\label{new}
     \mathcal{A}_l(\tau)= 3 \int_{0}^{\beta} \mathrm{d}t \ T_{i j k }(t) G_{l i}(\tau, t) G_{j k}(t, t),
\end{equation}
where the factor of $3$ comes from the multiplicity of the contraction. Let us now apply a useful technique: we act with the quadratic operator \( L_{ij}(\tau) = P_0 \frac{\mathrm{d}}{\mathrm{d} \tau} + P(\tau) \) (as defined in \eqref{L_operator}) to both sides of equation \eqref{new}. Recalling the definition of the Green's function from \eqref{verde}, this leads to the following system of coupled differential equations:
\begin{equation}\label{system2}
    \begin{cases}
        2 P_{p p}(\tau) \mathcal{A}_p (\tau) + 2 P_{p \varphi}(\tau) \mathcal{A}_{\varphi}(\tau) + \mathrm{i} \dot{\mathcal{A}_{\varphi}}(\tau)= 3 T_{p j k}(\tau) G_{j k}(\tau, \tau) \\
        2 P_{\varphi p}(\tau) \mathcal{A}_{p} (\tau)  + 2P_{\varphi \varphi}(\tau) \mathcal{A}_{\varphi}(\tau) - i \dot{\mathcal{A}_p}(\tau) = 3 T_{\varphi j k}(\tau) G_{j k}(\tau, \tau)
    \end{cases}.
\end{equation}
Solving for $\mathcal{A}_p(t)$ from the first equation and plugging it in the second equation, we end up with:
\begin{equation}\label{eq}
\begin{split}
    -\frac{\ddot{\mathcal{A}}_{\varphi}(\tau)}{2 P_{pp}}+\mathcal{A}_{\varphi}(\tau)\left(2 P_{\varphi \varphi}+i \frac{\dot{P}_{p \varphi}}{P_{pp}}-\frac{2\left(P_{p \varphi}\right)^2}{P_{pp}} \right)&=3 T_{\varphi j k}(\tau) G_{j k}(\tau, \tau)+ \\
    &+ \left(\frac{3 \mathrm{i} }{2 P_{pp}} \frac{\mathrm{d}}{\mathrm{d}\tau} -\frac{3 P_{p\varphi}}{P_{pp}}\right)T_{p j k}(\tau) G_{j k}(\tau, \tau).
\end{split}
\end{equation}
We now focus on the left-hand side of the equation and substitute the explicit forms of 
\(P_{pp}\), \(P_{\varphi \varphi}\), and \(P_{\varphi p}\), given in \eqref{PPPP}. After some straightforward algebraic manipulations, we arrive at the expression:
\begin{equation}
    -\frac{ \sin{(\theta)^2} }{8 P_{pp}} \left( \frac{\mathrm{d}^2}{\mathrm{d} \zeta^2} - \frac{2}{\cos{(\zeta)^2}} \right) \mathcal{A}_{\varphi}(\zeta) = \dots
\end{equation}
where we introduced the same $\zeta$ variable as before. The analysis of the right-hand side is more involved and requires nontrivial identities among the Green's functions. In particular, we need an expression for \( \frac{d}{d \tau} G(\tau, \tau) \) in terms of the Green's function itself. This derivation is carried out in detail in Appendix \ref{verdeverde}, and we present here the final result from \eqref{result}, written in component form \footnote{One can also explicitly verify that the components of the Green's function \eqref{ggg}, expressed in terms of the retarded and advanced Green's functions \eqref{complete_form1} and \eqref{complete_form2}, indeed satisfy the relations in \eqref{relations}.
 }:
\begin{equation}\label{relations}
\left\{\begin{array}{l}
 \frac{d}{d \tau} G_{pp}(\tau, \tau) =-4 \mathrm{i}\left(G_{pp}(\tau, \tau) P_{p \varphi}(\tau)+G_{p \varphi} (\tau, \tau)P_{\varphi \varphi}(\tau)\right) \\
 \frac{d}{d \tau} G_{p \varphi} (\tau, \tau) =2 \mathrm{i}\left(G_{pp}(\tau, \tau) P_{pp}(\tau)-G_{\varphi \varphi}(\tau, \tau)P_{\varphi \varphi}(\tau)\right) \\
 \frac{d}{d \tau} G_{\varphi \varphi} (\tau, \tau) =4 \mathrm{i}\left(G_{p \varphi}(\tau, \tau) P_{pp}(\tau)+G_{\varphi \varphi} (\tau, \tau)P_{p \varphi}(\tau)\right)
\end{array}\right.
\end{equation}
Using these relations, we can express the right-hand side of \eqref{eq} entirely in terms of the Green function components. This yields:
\begin{equation}
     \left( \frac{\mathrm{d}^2}{\mathrm{d} \zeta^2} - \frac{2}{\cos{(\zeta)^2}} \right) \mathcal{A}_{\varphi}(\zeta) = -\frac{24 P_{pp}}{\sin{(\theta)^2}} \left( c_{pp}(\tau) G_{pp}(\tau, \tau)+c_{p \varphi}(\tau) G_{p \varphi}(\tau, \tau) + c_{\varphi \varphi}(\tau) G_{\varphi \varphi}(\tau, \tau)\right)
\end{equation}
where the coefficients are:
\begin{equation}
\begin{aligned}
c_{\varphi\varphi}(\tau) & =T_{\varphi \varphi \varphi}+\frac{\mathrm{i}}{2 P_{pp}} \dot{T}_{p \varphi \varphi}-\frac{P_{\varphi p} T_{p \varphi \varphi}}{P_{pp}} + \frac{2T_{pp \varphi} P_{\varphi \varphi}}{P_{pp}} - \frac{2 T_{p \varphi \varphi} P_{p \varphi}}{P_{pp}} =  -\frac{\sin (\theta)^2}{24 P_{pp }} \frac{1}{ \sin \left(\theta+\frac{\tau}{2} \sin \theta\right)^2} \\
c_{p \varphi}(\tau) & = -2 \frac{P_{\varphi p} T_{pp\varphi}}{P_{pp}} +\frac{\mathrm{i}}{ P_{pp}} \dot{ T}_{p p \varphi}+\frac{2 T_{ppp} P_{\varphi \varphi}}{ P_{pp}}  =0 \\
c_{pp}(\tau) & =T_{\varphi pp}+\frac{\mathrm{i}}{2 P_{pp}} \dot{T}_{ppp}-\frac{P_{\varphi p}}{P_{pp}} T_{pp \phi}+\frac{2 T_{ppp}P_{p \varphi}}{ P_{pp}} -\frac{2 T_{pp \varphi} P_{pp}}{ P_{pp}} = 0 
\end{aligned}
\end{equation}
and are computed using \eqref{PPPP} and \eqref{T}.
Thus, we finally arrive at the following differential equation obeyed by $\mathcal{A}_{\varphi}$ \footnote{The same differential equation was also found in \cite{Okuyama:2023bch} in the context of the bilocal Liouville action derived from the $G\Sigma$ formulation of 
SYK.}:
\begin{equation}\label{diffe}
    \left( \frac{\mathrm{d}^2}{\mathrm{d} \zeta^2} - \frac{2}{\cos{(\zeta)^2}} \right) \mathcal{A}_{\varphi}(\zeta) = \frac{G_{\varphi \varphi}(\tau(\zeta), \tau(\zeta))}{\cos{(\zeta)^2}},
\end{equation}
To summarize, from \eqref{3.12} the one-loop correction to the correlator takes the form  
\begin{equation}\label{3.12}
\begin{split}
\langle e^{-2\Delta \varphi} \rangle_{\beta} &\simeq e^{-2\Delta \varphi_{\text{cl}}(\tau)} \left[ 1 - 2 \abs{\log q} \left( \Delta^2 G_{\varphi \varphi} + \Delta \mathcal{A}_{\varphi} \right) +\mathcal{O}\left(\abs{\log q}^2\right)\right],
\end{split}
\end{equation}
By comparing with \eqref{compa}, we observe that \( G_{\varphi \varphi} = -\mathcal{I} \) and \( \mathcal{A}_{\varphi} = -\mathcal{A} \). Indeed, substituting the expression for \( G_{\varphi \varphi} \) from \eqref{GG} and \( \mathcal{A}_{\varphi} = -\mathcal{A} \) from \eqref{one2} into \eqref{diffe}, we verify that the differential equation is satisfied. Furthermore, since \( \mathcal{A}_{\varphi}(\tau = 0) = \mathcal{A}_{\varphi}(\tau = \beta) = 0 \), due to the boundary conditions imposed on \( \delta \varphi \), the solution \( \mathcal{A}_{\varphi} = -\mathcal{A} \) to \eqref{diffe} is unique. This completes the proof and demonstrates perfect agreement between the one-loop gravitational correction to the bilocal operator and the one-loop correction in DSSYK.

\section{Conclusions}\label{5}
In this paper we have explicitly checked the duality between sine-dilaton gravity and DSSYK model, performing at one-loop order an honest path integral computation of the free energy and of the two-point fucntion of bilocal operators. We used the description of sine-dilaton gravity in terms of a deformed version of the standard Schwarzian model dual to JT gravity,  the so called q-Schwarzian theory. Particular attention has been devoted to the boundary condition of the one-loop determinant, crucially related to the correct Hartle-Hawking vacuum for the gravitational theory. Technically we have exploited a generalization of the Gel'fand-Yaglom theorem, adapting the standard formalism to our particular cases. We have found a small discrepancy between our result and previous computations \cite{Goel:2023svz, Okuyama:2023bch} for the partition function, that nevertheless does not affect the the two-point correlator. We attribute the discrepancy to a nontrivial ordering of the operators in the gravitational Hamiltonian that is involved during the quantization process. 

An interesting question regards the physical meaning of the one-loop corrections: one expects that this contribution should be related to the so called large $p$ limit of SYK, where $p$ is taken to be large, but
independent of $N$. 
We already observed that the q-Liouville field $\varphi$ represents conformal factor for the effective AdS$_2$ metric \eqref{sol}. At the leading order, this just describes different finite patches of the geometry as a function of $\beta$.  In \cite{Goel:2023svz}, the quantum correction $\delta \varphi$ was computed using the semiclassical saddle-point expansion of the DSSYK two-point function. This allowed for the calculation of the first quantum correction to the scalar curvature in the putative bulk dual geometry via $R=2 e^{2\varphi}\nabla^2 \varphi$. From this result, it was observed  that, as we move inside the bulk increasing $\tau$, the curvature significantly deviates from the the AdS$_2$ semiclassical saddle, indicating that the semiclassical approximation breaks down and the gravity theory becomes strongly coupled. Our computation of the boundary to boundary propagator of a matter field in sine-dilaton gravity, leading to the same result $\delta \varphi$ for the effective geodesic length measured by the probe \footnote{The quantum correction $\delta \varphi$ can just be found for instance by taking the derivative with respect to $\Delta$ of the result \eqref{3.12}.}, represents indeed a direct bulk derivation of the observation of \cite{Goel:2023svz}. As a result, we expect sine-dilaton gravity to also become strongly coupled as we move deeper into the bulk, rendering the semiclassical small-\(\abs{\log q}\) expansion unreliable in this regime. Nevertheless, the expansion may still be understood as an asymptotic series. By applying techniques such as resurgence theory, it might be possible to identify instantons, which could provide insights into the strongly coupled regime of sine-dilaton gravity. At the level of the partition function, the existence of such instantonic saddles has indeed been discussed in \cite{Blommaert:2024whf}.

As already remarked in the main text, the partition function of JT gravity can be exactly computed through equivariant localization \cite{Stanford:2017thb} (see \cite{Griguolo:2023aem} for an approach through supersymmetric localization), meaning that the one-loop approximation is basically exact. The same seems not to be true for the q-Schwarzian, where a full perturbative series in the coupling constant $\abs{\log q}$ is generated. One possibility is that a localization procedure could exploit the full geometrical structure of the Poisson sigma model or some supersymmetric generalization of it, involving directly the quantum group symmetry. A somehow related question concerns the characterization of the bilocal operator from a bulk point of view: in the gauge theory formulation in JT gravity they appear naturally as anchored Wilson lines \cite{Blommaert:2018oro, Iliesiu:2019xuh} therefore one suspects that the same role could be played by a ``quantum" holonomy in the Poisson sigma model.

There is also a certain number of direct extensions of the present paper that should be doable in near future. In JT gravity it is possible to implement twisted boundary conditions in the Schwarzian  in order to study the trumpet partition function and bilocal correlator on the trumpet \cite{Mertens:2019tcm}: it would be nice to engineer a similar process here for the q-Schwarzian, capturing the insertion of a defect in geometry of sine-dilaton gravity. On the other hand, the semiclassical expansion around classical solutions already suggests the presence of a geometrical defect to explain the restoration of the real periodicity in the model \cite{Blommaert:2024ydx}. A more comprehensive study of the trumpet geometries from the bulk point of view will be presented in \cite{newpaper3}.

One would like also to understand better the out-of-time-order correlator in the q-Schwarzian, starting from shock waves processes in sine-dilaton gravity, as originally done in JT gravity \cite{Lam:2018pvp}. Recently there have been also interesting proposals to connect DSSYK with three dimensional physics and de Sitter gravity \cite{Narovlansky:2023lfz, Verlinde:2024znh, Verlinde:2024zrh} and with higher dimensional supersymmetric theories \cite{Gaiotto:2024kze}: some non-trivial features of sine-dilaton gravity could be encoded there.

\newpage 

\section*{Acknowledgements}
We thank Andreas Blommaert, Carlo Meneghelli and Thomas Mertens for useful discussion and suggestions. LB, LG, LR, DS have been supported in part by the Italian Ministero dell’Università e della
Ricerca (MIUR), and by Istituto Nazionale di Fisica Nucleare (INFN) through the “Gauge and String Theory” (GAST) research project. JP acknowledges financial support from the European Research Council (grant BHHQG-101040024). Funded by the European Union. Views and opinions expressed are however those of the author(s) only and do not necessarily reflect those of the European Union or the European Research Council. Neither the European Union nor the granting authority can be held responsible for them.

\appendix
\newpage
\section{Green's functions}\label{general}
In this appendix, we want to show that it is possible to obtain Green's functions of a linear differential system if the fundamental solution of the associated homogeneous  problem is known. For clarity, we will firstly consider a differential operator that is diagonal in the derivatives (unlike the one we studied in the paper). As we shall see, however, it is straightforward to extend the methodology to off-diagonal operators. For more, see \cite{Dominici}, from which part of this appendix was inspired.
\paragraph{Diagonal operators.}Let $A(t)$ be a matrix of dimension $n \times n$ continuous for $t \in(\alpha, \beta)$ and let $a(t)$ be a vector of dimension $n \times 1$, continuous in the same interval. Let $M, N$ be constant matrices of dimension $n \times n$. We will be interested in finding the matrix Green's function of the following differential system (with homogeneous bilocal boundary conditions in $ t_1, t_2 \in(\alpha, \beta)$):
\begin{equation} \label{Non_Homogeneus_Diff}
	\left\{\begin{array}{l}
		\dot{x}(t)=A(t) x(t)+a(t) \\
		M x\left(t_1\right)+N x\left(t_2\right)=0
	\end{array}\right.
\end{equation}
Let's assume we know the fundamental (matrix) solution $K(t)$ of the associated homogeneous problem. By definition, then:
\begin{equation}
		\dot{K}(t)=A(t) K(t)
\end{equation}
Moreover, it is useful to introduce the following matrix, which encodes information about the homogeneous problem and the boundary conditions:
\begin{equation}
	\mathcal{R}(t_1,t_2) := \left[M K\left(t_1\right)+N K\left(t_2\right)\right]
\end{equation}
If $\det \mathcal{R} \neq 0$ and $\det K \neq 0$, it is always possible to define the following auxiliary bilocal function $\chi$:
\begin{equation}\label{Definition_Chi}
	\chi(t_1, t_2) :=-\mathcal{R}^{-1} N K\left(t_2\right) \int_{t_1}^{t_2} K^{-1}(\tau)\: a(\tau) d \tau
\end{equation}
A first fundamental statement is that a solution $x(t)$ of the non-homogeneous problem \eqref{Non_Homogeneus_Diff} can be written in terms of $\chi$ and $K(t)$, as follows:
\begin{equation}\label{x_solution}
	x(t)=K(t) \: \chi(t_1, t_2)+K(t) \int_{t_1}^t K^{-1}(\tau) a(\tau) d \tau
\end{equation}
We can prove this result simply by applying the definitions, indeed:
\begin{equation}\begin{split}
	&\dot{x}(t)=\dot{K}(t)\left[\chi+\int_{t_1}^t K^{-1}(\tau) a(\tau) d \tau\right]+K(t) K^{-1}(t) a(t) = 
	\\& = A(t) K(t)\left[\chi+\int_{t_1}^t K^{-1}(\tau) a(\tau) d \tau\right]+ a(t) =A(t) x(t)+a(t)
	\end{split}
\end{equation}
Furthermore, also the boundary conditions are satisfied\footnote{We notice that we can also express \eqref{Definition_Chi} as: 
$$
\left[M K\left(t_1\right)+N K\left(t_2\right)\right] \chi=-N K\left(t_2\right) \int_{t_1}^{t_2} K^{-1}(\tau) a(\tau) d \tau
$$}:
\begin{equation}
	M x\left(t_1\right)+N x\left(t_2\right)=M K\left(t_1\right) \chi+N\left[ K\left(t_2\right) \chi+ K\left(t_2\right) \int_{t_1}^{t_2} K^{-1}(\tau) a(\tau) d \tau \right]=0
\end{equation}
Now, for the purpose of extracting Green's functions, we notice that \eqref{x_solution} can be rewritten as an integral kernel, indeed:
\begin{equation}\label{x_solution_2}
	\begin{aligned}
		& x(t)=  K(t)\left(\chi+\int_{t_1}^t K^{-1}(\tau) a(\tau) d \tau\right) = \\
		= & K(t)\left[-\mathcal{R}^{-1} N K\left(t_2\right) \int_{t_1}^{t_2} K^{-1}(\tau) a(\tau) d \tau+\int_{t_1}^{t_2} K^{-1}(\tau) a(\tau) d \tau\right] =\\
		= & -K(t) \mathcal{R}^{-1} N K\left(t_2\right) \int_{t_1}^t K^{-1}(\tau) a(\tau) d \tau-K(t) \mathcal{R}^{-1} N K\left(t_2\right) \int_t^{t_2} K^{-1}(\tau) a(\tau) d \tau +\\
		& +K(t) \int_{t_1}^t K^{-1}(\tau) a(\tau) d \tau =\\
		= & \int_{t_1}^t\left[-K(t)\left(\mathcal{R}^{-1} N K\left(t_2\right)-1\right) K^{-1}(\tau)\right] a(\tau) d \tau  +\int_t^{t_2}\left[-K(t) \mathcal{R}^{-1} N K\left(t_2\right) K^{-1}(\tau)\right] a(\tau) d \tau
	\end{aligned}
\end{equation}
Introducing the Heaviside step function $\theta$, we can now define:
\begin{equation}\label{green}
	\begin{gathered}
		\tilde{G}(t, \tau) := \theta(t-\tau) \:\tilde{G}^{(a)}(t, \tau) + \theta(\tau - t) \: \tilde{G}^{(r)}(t, \tau)\\\\
		\tilde{G}^{(a)}(t, \tau):=-K(t)\left[\mathcal{R}^{-1} N K\left(t_2\right)-1\right] K^{-1}(\tau) \quad \text { if } \quad \tau<t \\\\
		\tilde{G}^{(r)}(t, \tau):=-K(t) \mathcal{R}^{-1} N K\left(t_2\right) K^{-1}(\tau) \quad \text { if } \quad \tau>t
	\end{gathered}
\end{equation}
and write \eqref{x_solution_2} as:
\begin{equation}\label{x_solution_3}
	x(t)=\int_{t_1}^{t_2} \tilde{G}(t, \tau) a(\tau) d \tau
\end{equation}
The last point is to show that $\tilde{G}(t, \tau)$ is indeed a Green's function associated to the differential operator of our interest $\tilde{L}(t) = \mathds{1} \: d/d t -  A(t)$. By simply applying this operator to \eqref{x_solution_3}, we get:
\begin{equation}
	\begin{gathered}
		\tilde{L}(t) x(t)=\tilde{L}(t) \int_{t_1}^{t_2} \tilde{G}(t, \tau) a(\tau) d \tau \\
		a(t)= \int_{t_1}^{t_2} \tilde{L}(t) \tilde{G}(t, \tau) a(\tau) d \tau
	\end{gathered}
\end{equation}
Hence, the following identity holds:
\begin{equation} \label{tilde_L}
	\tilde{L}(t) \tilde{G}(t, \tau) = \delta (t - \tau) \mathds{1}
\end{equation}
which is the defining property of a (matrix) Green's function.
\paragraph{Off-diagonal operators.}
As mentioned above, for the purposes of our paper, we want to search for the Green's function $G(t, \tau)$ of a more general differential operator:
\begin{equation}\label{L_operator_2}
	{L}(t) = P_0 \tilde{L}(t) 
\end{equation}
where $P_0$ is a constant, $n \times n$ and invertible matrix.
As a first step, we notice that the fundamental matrix of $L$ (denoted in the main text as $H(\tau)$) is equivalent to $K(\tau)$, the fundamental matrix of $\tilde{L}$. Since the two systems are in some way related, we want to find the link between the corresponding Green's functions. By inversion of \eqref{L_operator_2} and using \eqref{tilde_L}, we have:
\begin{equation}
	\begin{gathered}
	P_0^{-1} L(t) \tilde{G}(t, \tau) = \delta (t - \tau) \mathds{1}\\
	\Downarrow\\
	L(t) \tilde{G}(t, \tau) = \delta (t - \tau) P_0\\
	\Downarrow\\
	L(t) \tilde{G}(t, \tau) P_0^{-1} = \delta (t - \tau) \mathds{1}\\
\end{gathered}
\end{equation}
Hence we recognize the relation: 
\begin{equation}\label{green_2}
	G(t, \tau) = \tilde{G}(t, \tau) P_0^{-1}
\end{equation}

\paragraph{Explicit evaluation of Green's functions.}
We now turn our attention to the problem of finding the Green's function of the operator $L(t)$ in \eqref{L_operator} of the main text.
As observed in section \ref{dete}, the fundamental matrix of the system $H(\tau)$, presents a divergence when evaluated on the value $\tau= \beta$. Therefore, once again, we need to regularize it by introducing a cut-off $\epsilon$ and studying the problem in the interval $[0, \beta_{\epsilon}]$, where: $$\beta_{\epsilon}=  \frac{2 \pi -\epsilon -  4 \theta}{\sin(\theta)}$$This time the divergences coming from $H(\beta_{\epsilon})$ are exactly canceled by the contributions coming from the matrix $\mathcal{R}^{-1}$ and using the formulas \eqref{green} and \eqref{green_2} we can easily obtain the expressions for the Green's functions:
\begin{equation}\label{complete_form1}
	\begin{split}
		G^{(a)}_{\varphi \varphi}(\zeta, \upsilon)& = \mbox{ \small $ \frac{(\tan (\zeta )+\tan (\zeta ) (\zeta +u) \tan (u)+\tan (u)) (\tan (u) ((u-\upsilon ) \tan (\upsilon )-1)+\tan (\upsilon ))}{u \tan (u)+1}$ } \\
		G^{(a)}_{ \varphi p}(\zeta, \upsilon) &= \mbox{\small $ -\frac{\mathrm{i} \sec (\upsilon ) \csc (u+\upsilon ) (2 (u-\upsilon ) \sin (u)+\cos (u+2 \upsilon )+3 \cos (u)) (\tan (\zeta ) (\zeta +u) \sin (u)+\tan (\zeta ) \cos (u)+\sin (u))}{2 (u \sin (u)+\cos (u))}$} \\
		G^{(a)}_{p \varphi }(\zeta, \upsilon)&= \mbox{ \small $-\frac{\mathrm{i} \csc (\zeta +u) (\sin (\zeta ) (\tan (\zeta )+\tan (u))+\sec (\zeta ) (\zeta +u) \tan (u)) (\sin (u) ((u-\upsilon ) \tan (\upsilon )-1)+\cos (u) \tan (\upsilon ))}{u \tan (u)+1}$}\\
		G^{(a)}_{p p}(\zeta, \upsilon)&= \mbox{\small $-\frac{\sec (\zeta ) \cos (u) \sec (\upsilon ) \csc (u+\upsilon ) (\sin (\zeta )+(\zeta +u) \sin (u) \csc (\zeta +u)) (2 (u-\upsilon ) \sin (u)+\cos (u+2 \upsilon )+3 \cos (u))}{2 (u \sin (u)+\cos (u))}$}
\end{split}\end{equation}

\begin{equation}\label{complete_form2}
	\begin{split}
		G^{(r)}_{\varphi \varphi}(\zeta, \upsilon)& = \mbox{ \small $\frac{(\tan (\zeta )+\tan (u) (\tan (\zeta ) (u-\zeta )-1)) ((u+\upsilon ) \tan (\upsilon ) \tan (u)+\tan (u)+\tan (\upsilon ))}{u \tan (u)+1}$ } \\
		G^{(r)}_{ \varphi p}(\zeta, \upsilon) &= \mbox{\small $-\frac{\mathrm{i} \csc (u+\upsilon ) (\sin (u) (\tan (\zeta ) (u-\zeta )-1)+\tan (\zeta ) \cos (u)) (\sin (\upsilon ) (\tan (u)+\tan (\upsilon ))+(u+\upsilon ) \tan (u) \sec (\upsilon ))}{u \tan (u)+1}$} \\
		G^{(r)}_{p \varphi }(\zeta, \upsilon)&= \mbox{ \small $-\frac{\mathrm{i} \sec (\zeta ) \csc (\zeta +u) ((u+\upsilon ) \tan (\upsilon ) \tan (u)+\tan (u)+\tan (\upsilon )) (2 (u-\zeta ) \sin (u)+\cos (2 \zeta +u)+3 \cos (u))}{2 u \tan (u)+2}$} \\
		G^{(r)}_{p p}(\zeta, \upsilon)&= \mbox{\small $-\frac{\sec (\zeta ) \cos (u) \sec (\upsilon ) \csc (\zeta +u) (2 (u-\zeta ) \sin (u)+\cos (2 \zeta +u)+3 \cos (u)) ((u+\upsilon ) \sin (u) \csc (u+\upsilon )+\sin (\upsilon ))}{2 (u \sin (u)+\cos (u))}$}
\end{split}\end{equation}
In order to write the expressions above we only retained the non-vanishing terms of the $\epsilon$-series expansion; we also introduced $\zeta:=\frac{\pi}{2}-\theta-\frac{\tau}{2} \sin (\theta)$ and $u:=\frac{\pi}{2}-\theta$ as in \ref{semi}. From \eqref{complete_form1} and \eqref{complete_form2} it is then simple to obtain the complete Green's function by means of the definition $G(t, \tau) = \theta(t-\tau) \: G^{(a)}(t, \tau) + \theta(\tau - t) \: G^{(r)}(t, \tau)$.

\paragraph{Derivatives of Green's functions.}\label{verdeverde}
Here we want to find an expression for $\frac{d}{d t} G(t, t)$ in terms of the Green's function itself. By definition we have that:
\begin{equation}
 \frac{d}{d \tau} G(\tau, \tau)=\frac{d}{d \tau}\left(\frac{G^{(a)}(\tau, \tau)+G^{(r)}(\tau, \tau)}{2}\right)
\end{equation}
and for both the advanced and the retarded Green's function the following holds:
\begin{equation}
    \frac{d}{d \tau} G^{(i)}(\tau, \tau)=\left.\left(\frac{\partial }{\partial x} G^{(i)}(x, y) +\frac{\partial }{\partial y}G^{(i)}(x, y)\right)\right|_{x=y=\tau}.
\end{equation}
For the partial derivative with respect to the first argument we know that the Green's function has to satisfy:
\begin{equation}
 \frac{\partial}{\partial x}  G^{(a)}(x, y)=-P_0^{- 1} P(x) G^{(a)}(x, y)+\frac{1}{2} P_0^{-1} \delta(x, y)
\end{equation}
while by also observing that $G^{(a)}(x, y)=G^{(r)}(y, x)^{\top}$ we find the following relation:
\begin{equation}
\frac{\partial }{\partial y} G^{(a)}(x, y)=\frac{\partial }{\partial y} G^{(r)} \left( y, x \right)^{\top} = -G^{(r)}(y, x)^{\top} P(y)^{\top} \left(P_0^{-1}\right)^{\top}+ \frac{1}{2} \left(P_0^{-1}\right)^{\top} \delta(x-y)
\end{equation}
At the end, putting everything together, and observing that $\left(P(x)^{-1}\right)^{\top}=P(x)$ and that $P_0^{-1}= - \left(P_0^{-1}\right)^{ \top}$, we end up with the result:
\begin{equation}
\frac{\partial }{\partial x} G^{(a)}(x, y) +\frac{\partial }{\partial y}G^{(a)}(x, y) = -P_0^{-1} P(x) G^{(a)}(x, y) + G^{(a)}(x, y) P(y) P_0^{-1}
\end{equation}
Therefore, the derivative we are interested in is evaluated as:
\begin{equation}
    \frac{d}{d \tau} G^{(a)}(\tau, \tau) = -P_0^{-1} P(\tau) G^{(a)}(\tau, \tau) + G^{(a)}(\tau, \tau) P(\tau) P_0^{-1}
\end{equation}
Clearly the same is true for the retarded Green's function upon interchanging the indexes $a \leftrightarrow r$. Thus, upon summation, an analogous relation holds for the complete Green's function without indexes:
\begin{equation}
    \frac{d}{d \tau} G(\tau, \tau) = -P_0^{-1} P(\tau) G(\tau, \tau) + G(\tau, \tau) P(\tau) P_0^{-1}
\end{equation}
which in components and reads:
\begin{equation}\label{result}
\left\{\begin{array}{l}
 \frac{d}{d \tau} G_{pp}(\tau, \tau) =-4 i\left(G_{pp}(\tau, \tau) P_{p \phi}(\tau)+G_{p \phi} (\tau, \tau)P_{\phi \phi}(\tau)\right) \\
 \frac{d}{d \tau} G_{p \phi} (\tau, \tau) =2 i\left(G_{pp}(\tau, \tau) P_{pp}(\tau)-G_{\phi \phi}(\tau, \tau)P_{\phi \phi}(\tau)\right) \\
 \frac{d}{d \tau} G_{\phi \phi} (\tau, \tau) =4 i\left(G_{p \phi}(\tau, \tau) P_{pp}(\tau)+G_{\phi \phi} (\tau, \tau)P_{p \phi}(\tau)\right)
\end{array}\right.
\end{equation}

\section{Review: the Poisson sigma model formulation of sine-dilaton gravity}\label{Poisson}
In this section we elucidate a different perspective for the duality between  sine-dilaton gravity and the q-Liouville path integral \eqref{path1}, exploiting an intermediate description in terms of a Poisson sigma model \cite{Blommaert:2023opb}.
Let's start from the sine-dilaton action in
 \eqref{b+b} and let us introduce a zweibein one-form $e^{a}$, related to the metric via $g_{\mu \nu}=\eta_{ab}e^{a}_{\mu}e^{b}_{\nu}$, and the spin connection $\omega$, initially regarded as independent variables. 

 In this first order formalism, the sine-dilaton action \eqref{b+b} can then be rewritten as 
\begin{align}
    S=&\int\bigg( -\Phi \mathrm{d} \omega +\frac{\text{sin}(2 \abs{\log q} \Phi)}{2 \abs{\log q}} e^0\wedge e^1-\Phi_0(\mathrm{d} e^0-\omega\wedge e^1)-\Phi_1(\mathrm{d} e^1-\omega\wedge e^0)\bigg) \nonumber \\
     &+\int \Phi \omega+ S_{\mathrm{count}} \,,\label{first order}
\end{align}
where we denoted with $S_{\mathrm{count}}$ the boundary action
\begin{align}
    S_{\mathrm{count}}=\int \bigg( \Phi_0 e^{0}+\Phi_1 e^{1}-\mathrm{d} t  \ \mathbf{H}(\Phi_0,\Phi_1,\Phi) \bigg) \,,\label{contro}
\end{align}
and $\mathbf{H}(\Phi_0,\Phi_1,\Phi)$ is the boundary Hamiltonian, given by
 \begin{equation}\label{H}
     \mathbf{H}(\Phi_0,\Phi_1,\Phi)=-\Phi_{0}^{2}+\Phi_{1}^{2}-\frac{\cos\left(2 \abs{\log q} \Phi\right)}{2 \abs{\log q}}\,.
\end{equation}
To show this is just sine-dilaton gravity, one can integrate out $\Phi_0$ and $\Phi_1$ in \eqref{first order}, enforcing the torsion constraints $T^{a}=\mathrm{d} e^{a}+\epsilon^{a b} \omega \wedge e_{b}=0$, which allow to determine $\omega$ in terms of the zweibein. If we plug the solution back into the action, we transition to the second-order formulation and precisely recover the bulk part of the sine dilation action \eqref{b+b}, by further exploiting that $\frac12 \sqrt{-g} R \mathrm{d}^2 x=-\mathrm{d} \omega$ and that $e^0\wedge e^1$ is the two dimensional volume form. Moreover, the $\Phi \omega$ boundary term in \eqref{first order} yields the Gibbons-Hawking-York curvature term, while  $S_{\mathrm{count}}$ corresponds to the boundary counterterm introduced in \eqref{b+b}. To show this, as argued in \cite{Blommaert:2023wad}, we consider the solution of the equation of motion obtained by varying \eqref{first order} with respect to $\omega$, in the gauge where $\Phi_1=0$ and $\Phi=r$, i.e.
\begin{equation}\label{on-shell}
\Phi_0=\frac{1}{e^{1}_r}=F(r)^{1/2}
\end{equation}
Plugging this solution, with $F(r)$ given in \eqref{metrrr} , inside the Hamiltonian \eqref{H} exactly reproduces the ADM energy \eqref{energy} \footnote{Once we restore the correct $\abs{\log q}$ factors prior to rescaling of the dilaton.}.
On the other hand, since $e^{0}_{\tau}=F(r)^{1/2}$, the full boundary action $S_{\mathrm{count}}$ in \eqref{contro} reduces on shell to:
\begin{align}
    S_{\mathrm{count}}=\int \mathrm{d} t  \ \frac{\cos\left(2 \abs{\log q} \Phi\right)}{2 \abs{\log q}} \,,\label{contro2}
\end{align}
which exactly matches with the boundary counterterm in \eqref{b+b}, once the boundary conditions \eqref{bcrescaled} are imposed. This analysis thereby validates the form of the boundary terms in \eqref{first order} and the boundary Hamiltonian \eqref{H}. 

Having established that \eqref {first order} corresponds to sine-dilaton gravity in the first order formulation, we now proceed to show that \eqref {first order} can indeed be reformulated as a \textit{Poisson sigma model} \cite{Ikeda:1993aj,Ikeda:1993fh,Schaller:1994es}. This can be seen to the generalization of the first order reformulation of JT gravity as a BF gauge theory. To achieve this, we perform the following field redefinition:
\begin{equation}
    \Phi=-J_H\,,\quad \Phi_0=-J_1\,,\quad \Phi_1=-J_0\,,\quad \omega=A_H\,,\quad e^1=A_0\,,\quad e^0=A_1\,,\label{mappp}
\end{equation}
in terms of a gauge connection $(A_\mu)_A$ and a three dimensional coordinate $J_{A}$, with label $A=0,1,H$. Then, by introducing the notation \cite{Cattaneo:2001bp}
\begin{equation}
     \alpha_{H 0}=\left\{J_H,J_0\right\}=-J_1\,,\quad \alpha_{H 1}=\left\{J_H,J_1\right\}=-J_0\,,\quad \alpha_{0 1}=\left\{J_0,J_1\right\}=\frac{\sin(2\abs{\log q} J_H)}{2 \abs{\log q}}\,,\label{mappp2}
\end{equation}
the sine-dilaton gravity in the form \eqref{first order} gets rewritten as
\begin{align}\label{actionPSM}
     S_\text{PSM}=&\int\bigg(J_B\,\mathrm{d} A_B+\frac{1}{2}\alpha_{B C}(J_A)A_B\wedge A_C\bigg)- \int\bigg(J_B A_B
     +\mathrm{d} t\,  \mathbf{H}(J_A)\bigg)\nonumber\\
     =&\int_{u_1}^{u_2} \mathrm{d} u\int \mathrm{d} t\bigg(-(A_u)_B\dot{J_B}+(A_t)_B\bigg(J_B'+\left\{J_B,J_C\right\}\, (A_u)_C\bigg)\bigg)-\int\mathrm{d} t\, \mathbf{H}(J_A)\,,
\end{align}
where in the second step we specified our two-dimensional space to a strip $I \times \mathbb{R}$ with coordinates $(t,u)$ and $u \in \left[u_L,u_R\right]$, explicitly wrote the exterior derivative and integrated it by parts.

This is indeed a \textit{Poisson sigma model}, with the coordinates $J_{A}$ parametrizing a three-dimensional ``target space'' characterized by the Poisson structure \eqref{mappp2}, specific for our sine-dilaton model. Readers may recognize in it the classical Poisson bracket algebra that, upon quantization, reproduces the $SU_q(1,1)$ quantum group algebra. However, the Poisson algebra \eqref{mappp2} is here an an external structure of target space, while at the end we wish to identify the above with the symmetry algebra of the dynamical system itself, namely the q-Schwarzian living on the boundary. To achieve this, we will need to reduce the infinite dimensional phase space of the \textit{Poisson sigma model} to the 6-dimensional phase space characterizing the q-Schwarzian \cite{Blommaert:2023opb}.

One approach to determine the classical phase space of the 6d q-Schwarzian is detailed in \cite{Blommaert:2023wad}. Here, we offer a shorter and more heuristic argument to demonstrate that the appropriate algebra emerges on the boundary, giving rise to the dynamics of the q-Schwarzian. This is done by integrating out the Lagrange multiplier $(A_t)_B$ in \eqref{actionPSM}, which imposes the constraint  
 \begin{equation}\label{constraint}
     J^{'}_{A}=-\left\{J_A,J_B\right\}\, (A_u)^{B}\,.
\end{equation}
Define $j_{A}^{L}=J_{A}(u_{L})$ and $j_{A}^{R}=-J_{A}(u_{R})$, i.e. the value of the target coordinates at the endpoints of the interval.
 By integrating \eqref{constraint}, we are able to relate spatially separated values of $J_{A}$. In particular, starting from a generic bulk point $J_{A}(u)$ we can compute the boundary value $j_{A}^{L}$ as
 \begin{equation}\label{constraint2}
     j_{A}^{L}=J_{A}(u)+\int_{u_L}^{u} \mathrm{d} u_1  \ \alpha_{AC}\left(J(u_1)\right)(A_{u})^{C}(u_1)\,.
 \end{equation}
 Moreover, the first term in
 the \textit{Poisson sigma model} Lagrangian \eqref{actionPSM} tells us that $A_{u}$ and $J$ are canonically conjugated variables, so on a spacial slice they obey the Poisson brackets
 \begin{equation}\label{PB}
   \left\{(A_{u})_{B}(u_1),J_{C}(u_2)\right\}=\delta_{BC}\delta(u_1-u_2)\,.
\end{equation}
 By exploiting \eqref{constraint2} and \eqref{PB}, one can thus compute
 \begin{equation}
     \left\{j_{A}^{L},j_{B}^{L}\right\}=\int_{u_L}^{u} \mathrm{d} u_1  \ \alpha_{AC}\left(J(u_1)\right) \left\{(A_{u})^{C}(u_1),J_{B}(u_L)\right\}=\alpha_{AB}\left(j^{L}\right)\,.
 \end{equation}
 With a similar argument for the right boundary, one can show that the following Poisson brackets hold \cite{Cattaneo:2001bp}:
\begin{equation}\label{six}
    \left\{j_{A}^{L},j_{B}^{L}\right\}=\alpha_{AB}\left(j^{L}\right) \qquad \left\{j_{A}^{R},j_{B}^{R}\right\}=\alpha_{AB}\left(j^{R}\right) \qquad \left\{j_{A}^{R},j_{B}^{R}\right\}=0\,.
 \end{equation}
which indeed form a six-dimensional phase space variables and correspond to two copies of the current algebra
\begin{equation}
    \{h,e\}=e\,,\quad  \{h,f\}=-f\,,\quad  \{e,f\}=\frac{\sin(2\abs{\log q}h)}{\abs{\log q}}\,,\label{chargealgebra}
\end{equation}
once we define $-2j_0=e+f$, $2j_1=e-f$, $j_{h}=h$. We hence argue we can "diagonalize" this six-dimensional space \eqref{six} by
the phase space variables $\left(\varphi,\beta,\gamma,p_{\varphi},p_{\beta},p_{\gamma}\right)$ of the q-Schwarzian system, where the currents $j_{A}$ become functions of these canonical coordinates, such that $\left\{x_{A},p_{B}\right\}=\delta_{AB}$. Moreover, the boundary Hamiltonian \eqref{H} precisely corresponds to the q-Schwarzian Hamiltonian \cite{Blommaert:2023opb}.
We thus conclude the \textit{Poisson sigma model} path integral \eqref{actionPSM}, after solving the constraint, reduces on the boundary to the path integral of a particle on $SU_{q}(1, 1)$ \cite{Blommaert:2023opb}, which we report here:
\begin{equation}\label{q_schw}
\begin{split}
\int\mathcal{D}\varphi \mathcal{D}p_{\varphi} \mathcal{D}\beta \mathcal{D}p_{\beta}\mathcal{D}\gamma \mathcal{D}p_{\gamma} \,\, &\exp\left(\int \mathrm{d} t \left( \mathrm{i} p_{\varphi} \varphi' +\mathrm{i} p_{\beta} \beta'+\mathrm{i} p_{\gamma} \gamma'+ \frac{1}{2 \abs{\log q}} \cos(\abs{\log q}p_{\varphi}) \right.\right. \\
&\left.\left.+ \frac{1}{4 \abs{\log q}} e^{\mathrm{i} \abs{\log q} p_{\varphi}} e^{-2 \varphi} \frac{\left(e^{2 \mathrm{i} \abs{\log q}\beta p_{\beta}}-1\right)\left(e^{2 \mathrm{i} \abs{\log q}\gamma p_{\gamma}}-1\right)}{\beta \gamma}\right) \right)
\end{split}
\end{equation}
One can explicitly prove that the 6 conserved charges of this system exactly satisfy two copies of the algebra \eqref{chargealgebra}.In the case of JT gravity, one has to introduce some constraints on the $\mathfrak{sl}(2,\mathbb{R})$ generators to actually implement the Brown-Henneaux asymptotic boundary conditions; these are the so-called mixed parabolic boundary conditions. In \cite{Blommaert:2023opb} it is argued the correct generalization of those, in order to reduce \eqref{q_schw} to the DSSYK chord quantum mechanics, is 
\begin{equation}
    e^{-\mathrm{i}\abs{\log q}h} f=\frac{\mathrm{i}}{2\abs{\log q}}\,.\label{1.21 const}
\end{equation}
Specifically, will need to impose two such asymptotic constraints \eqref{1.21 const} to obtain the path integral description for a Cauchy slice with two asymptotic boundaries.
In doing so, the phase space reduces again from 6d to 2d, leading to the q-Liouville path integral \eqref{path1} in terms of $\varphi,p_{\varphi}$ fields only, since one can gauge fix $\beta=\gamma=0$.

Furthermore, because of the previous identifications and the solution \eqref{on-shell}, one has that
\begin{equation}
h=-\Phi_{\mathrm{bdy}} \qquad f=\Phi_{0,\mathrm{bdy}}=\sqrt{F}
\end{equation}
which indeed shows the mixed parabolic boundary conditions \eqref{1.21 const} correspond to the gravitational boundary conditions we imposed in \eqref{bcrescaled}.

\bibliographystyle{unsrt}
\bibliography{bibliography}

\end{document}